\documentclass[aps,prl,reprint,superscriptaddress,showpacs]{revtex4-1}
\usepackage{color}
\usepackage{graphicx}
\usepackage{dcolumn}
\usepackage{float}
\usepackage{graphicx}
\usepackage{amssymb}
\definecolor{olive}{RGB}{0,128,0}
\definecolor{magenta}{RGB}{255,0,255}
\definecolor{darkyellow}{RGB}{128,128,0}
\definecolor{wine}{RGB}{128,0,0}

\begin{document}


\title{\textbf{
The Quantified NTO Analysis for the Electronic Excitations of Molecular Many-Body Systems}}


\author{Jian-Hao Li}
\affiliation{Department of Physics, Center for Theoretical Sciences and Center for Quantum Science and Engineering, National Taiwan University, Taipei 10617, Taiwan}
\affiliation{Center for Condensed Matter Sciences, National Taiwan University, Taipei 10617, Taiwan}
\author{Jeng-Da Chai}
\affiliation{Department of Physics, Center for Theoretical Sciences and Center for Quantum Science and Engineering, National Taiwan University, Taipei 10617, Taiwan}
\author{Guang-Yu Guo}
\affiliation{Department of Physics, Center for Theoretical Sciences and Center for Quantum Science and Engineering, National Taiwan University, Taipei 10617, Taiwan}
\affiliation{Graduate Institute of Applied Physics, National Chengchi University, Taipei 11605, Taiwan}
\author{Michitoshi Hayashi}
\affiliation{Center for Condensed Matter Sciences, National Taiwan University, Taipei 10617, Taiwan}


\date{\today}

\begin{abstract}
We show that the origin of electronic transitions of molecular many-body systems can be revealed by a quantified
natural transition orbitals (QNTO) analysis and the electronic excitations of the total system can be
mapped onto a standard orbitals set of a reference system.  We further illustrate QNTO on molecular
systems by studying the origin of electronic transitions of DNA moiety, thymine and thymidine. This QNTO
analysis also allows us to assess the performance of various functionals used in time-dependent density functional response theory.
\end{abstract}

\pacs{31.10.+z, 31.15.ag, 31.15.ee}

\maketitle
Quantum mechanical description of electronic excitations of molecular many-particle systems has been
an important topic to date. Time-dependent linear response DFT (TDDFT)~\cite{R-TDDFT, R-LRTDDFT} has been widely used to calculate
the electronic excitation energies and oscillator strengths of large molecular systems.  In contrast, high-level quantum chemistry
approaches such as equation of motion coupled cluster singles and doubles (EOM-CCSD) ~\cite{R-EOM} can deal with only small systems.
The TDDFT results can be, however, unreliable due to the quality of exchange-correlation functionals used.
While assessment of different functionals often focuses on the excitation energies and oscillator strengths
~\cite{R-LC10, R-LC11}, verification of the physical origin of electronic transitions should be desirable.
Casida has proposed that TDDFT wavefunction can be interpreted in terms of linear combination of singly excited
configurations (LCSEC)~\cite{R-LRTDDFT}.  Casida's scheme suggests that a systematic analysis can be developed
so that LCSEC of TDDFT and many-body theories like EOM-CCSD can be linked for each excitation.
The LCSEC contains information on transition origin that directly connects to the electronic structure change
which governs the physics and chemical actions, like nuclear dynamics and chemical reactions, of an electronically excited molecule.

For assigning the transition origin of an excitation from LCSEC, the most important property is its universality.
In other words, each excitation can only have one unique representation of transition origin.
In this regard, we find that natural transition orbital (NTO) analysis [6] is a good starting point because it
relies on the minimum representation of the transition operator using a NTO pair consisting of an electron-
and a hole-orbital for an excitation.
However, we also find that in many molecular systems, an excitation cannot be described very well by single NTO pair.
One of the typical and simplest examples is a well separated dimer.
Therefore, we develop a quantified NTO (QNTO) analysis that allows taking into account multi-NTO pairs properly as needed.
Moreover, NTO analysis uses a pictorial description of a NTO pair that can result in ambiguity especially for large molecules
with no symmetry.  To establish a clear interpretation of transition origin of NTO pairs, we introduce a
standard-orbitals set onto which the hole- and electron-orbital of a NTO pair can be mapped.  This scheme can be applied
for a wide range of molecular systems, such as DNA and proteins.
With QNTO analysis, we demonstrate that separate dimer can be properly described by two NTO pairs.
In addition, the variation of the transition origin of an excitation due to the environment change of the system can also
be determined based on the coefficient change of standard-orbitals.
Furthermore, we show that TDDFT together with recently developed long-range corrected (LC) hybrid
functionals~\cite{R-LC1,R-LC2,R-LC3,R-LC4,R-LC5,R-LC6,R-LC9,R-LC10,R-LC11} (LC-TDDFT)
can predict electronic transition origins of similar quality as EOM-CCSD.
\setlength{\parskip}{0cm}

For single-reference linear response methods for electronic excitations, e.g., configuration interaction singles (CIS),
time-dependent Hartree-Fock (TD-HF), EOM-CC, and TDDFT, the NTO analysis~\cite{R-NTO} is known to give a
concise transition picture out of the original many SECs of an (often low-lying) excitation. 
This is due to the fact that mathematically some SECs in a LCSEC can be combined to form the minimum representation
of transition operator.  For example, let us consider an electronic excitation whose LCSEC is $0.3\Psi^{a}_{i}+0.5\Psi^{a}_{j}+0.8\Psi^{b}_{j}$
where, for example, $\Psi^{a}_{i}=\hat{a}^{\dag}_{a}\hat{a}_{i}\Psi$ and $\Psi$ is the ground state wavefunction of
a theoretical method which may be a single determinant (e.g. in DFT) or a combination of multi-determinants (e.g. in EOM-CCSD).
If one follows a NTO-generating procedure - a unitary transformation to the original occupied and virtual orbitals
- the best reduced LCSEC can be achieved with an additional phase confirmation of SECs.  For the above example, it reduces to $0.3\Psi^{a}_{i}+0.5\Psi^{a}_{j}+0.8\Psi^{b}_{j}=\sqrt{0.9172}\Psi^{A}_{I}+\sqrt{0.0628}\Psi^{B}_{J}$ where $\hat{a}_{I}=N_{I}(c_{1}\hat{a}_{i}+c_{2}\hat{a}_{j})$, $\hat{a}_{J}=N_{J}(-c_{2}\hat{a}_{i}+c_{1}\hat{a}_{j})$, $\hat{a}^{\dag}_{A}=N_{A}(d_{1}\hat{a}^{\dag}_{a}+d_{2}\hat{a}^{\dag}_{b})$ and $\hat{a}^{\dag}_{B}=N_{B}(-d_{2}\hat{a}^{\dag}_{a}+d_{1}\hat{a}^{\dag}_{b})$ with $c_{1}=0.5696$, $c_{2}=0.8219$, $d_{1}=0.1784$, $d_{2}=0.9840$, and, for instance, $N_{I}=1/\sqrt{c^{2}_{1}+c^{2}_{2}}$.  In this new description, the first SEC $\Psi^{A}_{I}$, corresponding to the first dominant NTO pair (NTO1), now dominates and therefore the transition picture of the excitation is clear.

However, as above mentioned, there may be a case in which the NTO1 cannot be representative to an excitation, e.g. in a separate dimer.
For this reason, in the first step, we generalize the traditional NTO analysis by taking into account the 2nd dominant NTO pair (NTO2)
if the NTO1 has less than 70\% domination of total LCSEC.  We also establish a procedure to determine the relative phase between NTO1 and NTO2.
Adding NTO3 or higher term can readily be done.  For the physical properties associated with one-particle operators like electronic density,
the phase difference leads to no difference, $\langle\Psi^{a}_{i}\pm\Psi^{b}_{j}|\hat{\rho}(\mathbf{r})|\Psi^{a}_{i}\pm\Psi^{b}_{j}\rangle=\langle\Psi^{a}_{i}|\hat{\rho}(\mathbf{r})|\Psi^{a}_{i}\rangle+\langle\Psi^{b}_{j}|\hat{\rho}(\mathbf{r})|\Psi^{b}_{j}\rangle$, $i \neq j$,
while it is not for two-particle operators such as the interelectron coulomb repulsion.

As the 2nd step of generalization, we project the hole- (NTO1-H) and electron-orbital (NTO1-E) of NTO1 onto a chosen standard orbitals set to see their quantitative involvement.  The same procedure will be performed for NTO2.  In doing so three advantages emerge:

(1) The precise contribution of a particular orbital-type of interest in an excitation can be computed.

We consider $0.3\Psi^{a}_{i}+0.5\Psi^{a}_{j}+0.8\Psi^{b}_{j}$ as an example whose NTO1 is $\Psi^{A}_{I}$,
if we simply use $\{\varphi_{i}, \varphi_{j}, \varphi_{a}, \varphi_{b}\}$ as the standard-orbitals,
the proportions of $\varphi_{i}$ and $\varphi_{j}$ in the NTO1-H are 0.1784 and 0.9840, respectively.
Similarly, the proportions of $\varphi_{a}$ and $\varphi_{b}$ in the NTO1-E are 0.5696 and 0.8219, respectively.
Having these results, say, if $\varphi_{i}$ and $\varphi_{j}$ originate from the hole-orbitals of $^{1}n\pi^{*}$
and $^{1}\pi\pi^{*}$ transitions, respectively, then the precise proportion of the two orbitals in the excitation
of interest can be obtained. Since NTO-based LCSEC is unique, the orbital projection would not be misleading.

(2) Improving the pictorial NTO interpretation.

Although NTO analysis can largely reduce many SECs in the original LCSEC to the dominant one or two,
it is only rendered as an assistant to see the qualitative transition picture by plotting NTO1-H and NTO1-E.
On the other hand, projecting NTO1(2)-H(E) to standard-orbitals set would provide a rigorous explanation of transition origin.

(3) Unification of orbital-standard for comparison.

If the exact ground state wavefunction of an electronic system can be approximated by either $|\Psi^{KS}_{0}\rangle$
of DFT or $|\Psi^{CCSD}_{0}\rangle$ of CCSD,
the wavefunction up to LCSEC for an electronic excitation can be obtained within
the linear response theory.
However, the LCSECs given by TDDFT and EOM-CCSD are based on
different references, thus making a direct comparison difficult. Furthermore, ambiguity can also arise
if one compares LCSECs for excitations in two systems with different environments.
Therefore, importantly, projection of NTO1(2)-H(E) of LCSEC onto a standard-orbitals would make it possible to compare
two LCSECs based on different references.

\begin{table}
\caption{
Recovery of exciton theory prediction (TD-$\omega$B97X) for (a)$^{1}\pi\pi^{*}$ and (b)$^{1}n\pi^{*}$ excitations.
'P' denotes the phase of NTO2.  In (a) as the C6-C5' distance, defined in ~\cite{R-Supp}, increases, the two
local transitions (NTO1, NTO2) tend to have a similar domination.  The phase flipping between 9.24{\AA} and 10.24{\AA}
may be due to a conical intersection.  In contrast, in (b) two local transitions remain uncoupled
regardless of the C6-C5' distance.}
\begin{tabular}{|c|cccc|c|cccc|c|}
\hline
  (a)   &          \multicolumn{ 4}{c|}{NTO1:"P1$_{v}$ - S1$_{v}$"} &          &          \multicolumn{ 4}{c|}{NTO1:"P1$_{u}$ - S1$_{u}$"} &           \\

       C6-C5' &          \multicolumn{ 4}{c|}{NTO2:"P1$_{u}$ - S1$_{u}$"} & P           &          \multicolumn{ 4}{c|}{NTO2:"P1$_{v}$ - S1$_{v}$"} & P           \\

   ({\AA}) & $\lambda$ &       f &   1st &   2nd &  &  $\lambda$ &      f &     1st &  2nd &          \\
           & (nm)      &         &   (\%) &   (\%) &  &      (nm) &         &     (\%) &  (\%) &          \\
\hline
      4.24 &   230.4   &  0.104  &     65 &     32 & - &    228.4  &  0.227  &     65 &     31 &        + \\
\hline
      5.24 &   229.6   &  0.114  &     69 &     28 & - &    228.4  &  0.242  &     69 &     28 &        + \\
\hline
      6.24 &   229.3   &  0.118  &     68 &     28 & - &    228.5  &  0.252  &     68 &     28 &        + \\
\hline
      7.24 &   229.2   &  0.119  &     65 &     32 & - &    228.7  &  0.261  &     65 &     32 &        + \\
\hline
      8.24 &   229.1   &  0.118  &     57 &     39 & - &    228.7  &  0.268  &     57 &     39 &        + \\
\hline
      9.24 &   229.0   &  0.118  &     50 &     46 & - &    228.8  &  0.272  &     50 &     46 &        + \\
\hline
     10.24 &   228.8   &  0.270  &     64 &     33 & + &    229.0  &  0.123  &     64 &     33 &        - \\
\hline
\end{tabular}

\raggedright
\begin{tabular}{|c|ccc|ccc|}
\hline
  (b)    &          \multicolumn{ 3}{c|}{NTO1:"N1$_{u}$ - S1$_{u}$"} &          \multicolumn{ 3}{c|}{NTO1:"N1$_{v}$ - S1$_{v}$"}\\

  C6-C5' & $\lambda$ &        f &  1st & $\lambda$ &        f &  1st  \\
 ({\AA}) &      (nm) &          &   (\%) &      (nm) &          &  (\%) \\
\hline
    4.24 &    243.9  &  0.0001  &     99 &   242.14  &  0.0002  &    99 \\
\hline
    7.24 &    242.8  &  0.0002  &    100 &   242.10  &  0.0002  &   100 \\
\hline
   10.24 &    242.5  &  0.0002  &    100 &   242.11  &  0.0002  &   100 \\
\hline
\end{tabular}
\end{table}

We demonstrate that QNTO analysis correctly recovers the molecular exciton theory~\cite{R-Exciton} prediction
in which a molecular homo dimer system composed of well separated geometrically identical two monomers (ideal dimer)
has the formation of two excited states, $\frac{1}{\sqrt{2}}(\varphi^{\dag}_{u}\varphi_{v}\pm\varphi_{u}\varphi^{\dag}_{v})$,
where $\varphi^{\dag}_{u}$ and $\varphi_{u}$ is the excited and unexcited wavefunctions of monomer u
and $\varphi^{\dag}_{v}$ and $\varphi_{v}$ those of monomer v.  This situation should be reflected in
QNTO analysis by having NTO1 and NTO2 as each monomer's local transition with close domination.
QNTO helps verify if a real dimer system can be appropriately described by the exciton model or not.
Table I shows this trend of two $^{1}\pi\pi^{*}$ excitations and no similar trend for two $^{1}n\pi^{*}$
ones of a real homo dimer system composed of two geometrically slightly different monomers in various
separations.  In addition, QNTO can be applied to molecular hetero dimer as well.

After generalizing NTO to QNTO analysis, obtaining a reliable LCSEC for an excited state is the
next important concern.  It depends on the use of proper approximate exchange-correlation functionals
in TDDFT calculations mentioned earlier. In this regard, recently developed LC hybrid functionals,
are proved to be good candidates to describe long-range transitions, and sometimes even outperform
the traditional (non-LC) hybrid functionals for short-range transition as well.  In a typical LC hybrid functional,
100\% Hartree-Fock exchange is employed for a long-range part of the interelectron repulsive operator L(r)/r,
while a GGA exchange functional is employed for the complementary short-range part. Recently, Chai and Head-Gordon
found that LC hybrid functionals can outperform global hybrid functionals (e.g. B3LYP) in several important applications,
such as thermochemistry, thermal kinetics, noncovalent systems, dissociation of symmetric radical cations
and long-range charge-transfer excitations between two well-separated molecules~\cite{R-LC3,R-LC4,R-LC5,R-LC6}.
Recent TDDFT performance tests of various LC- and non-LC hybrid functionals for predicting
excitation properties of several molecular systems have been reported. Jacquemin \emph{et al.} assessed
the accuracy of $\omega$B97 family for predicting excitation energies ~\cite{R-LC10}, whereas Caricato \emph{et al.} focused
on the simulated spectra of three molecular groups - Alkenes, Carbonyls and Azabenzenes - in reference to
EOM-CCSD results ~\cite{R-LC11}.  Since here we emphasize NTO1(2) transition origin of electronic excitations,
several excitations of two benchmark systems - thymine (Thy) and thymidine (dT) - calculated by various
theoretical methods are investigated as examples.

\begin{figure}
\raggedright
(a)
\includegraphics[angle=0,width=1.59cm]{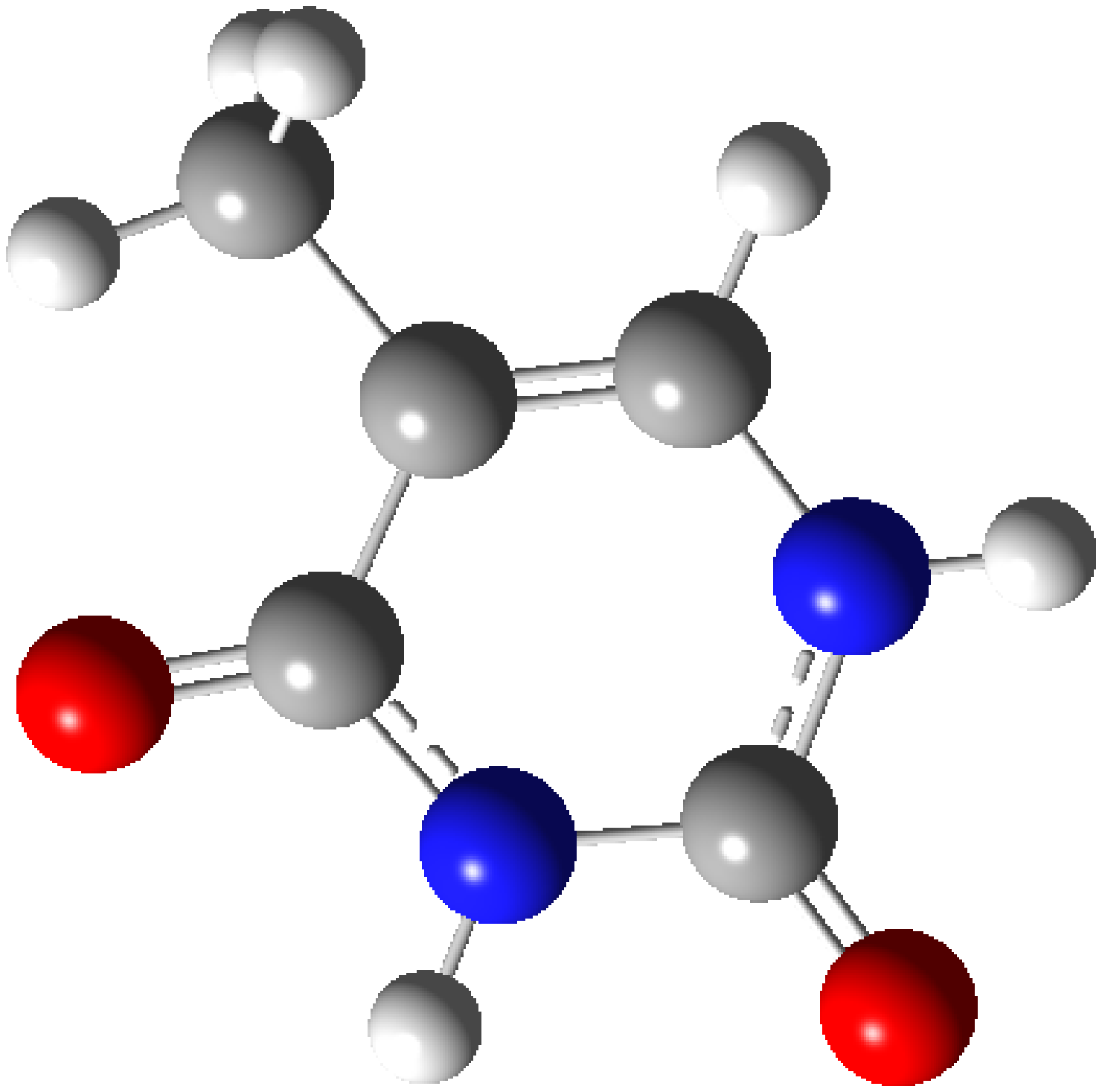}
(b)
\includegraphics[angle=0,width=2cm]{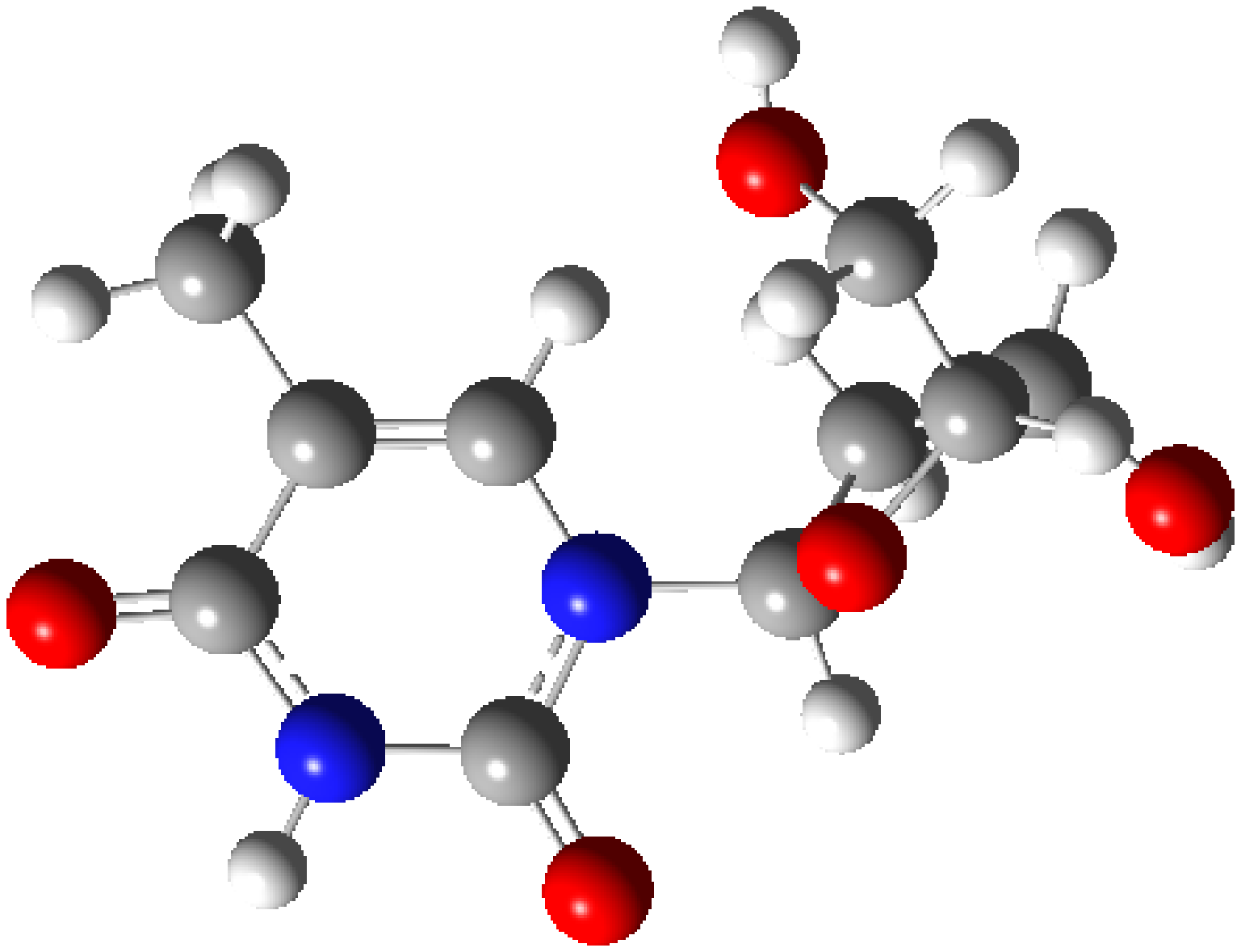}
\caption{Molecular structrues of (a) single thymine (Thy) and (b) thymidine (dT) extracted from ideal B-DNA.  The dT has an additional backbone structure of sugar connecting to Thy.}
\end{figure}

The prepared Thy and dT (Fig. 1) are extracted from ideal B-DNA structure.  EOM-CCSD and TDDFT singlet excitations
calculations are carried out using Gaussian09 package~\cite{R-G09} with 6-31G(d) basis set.  For simplicity,
solvent models are not used. EOM-CCSD and
TD-$\omega$B97X~\cite{R-LC3} are demonstrated here whereas several other theoretical methods are tested in ~\cite{R-Supp}.


\begin{figure}
(a)
\includegraphics[angle=0,width=1.1cm]{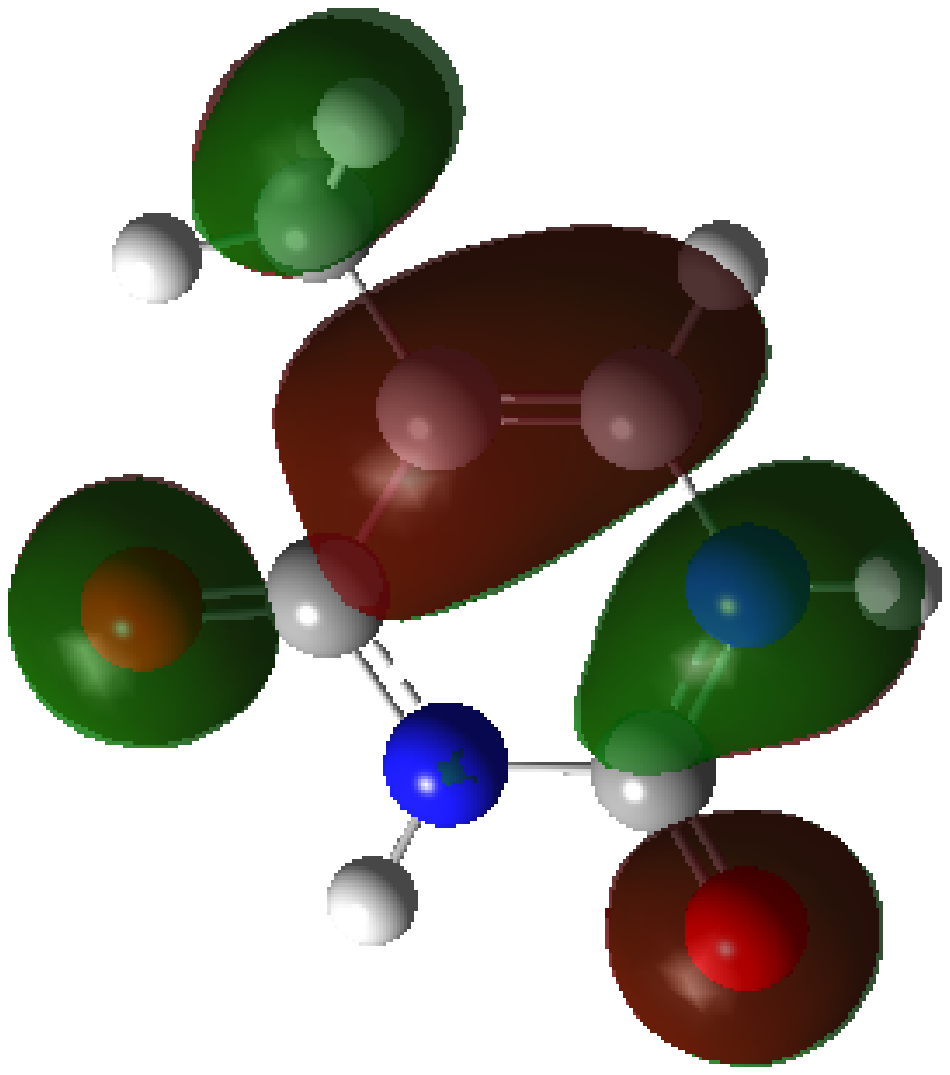}
\includegraphics[angle=0,width=1.1cm]{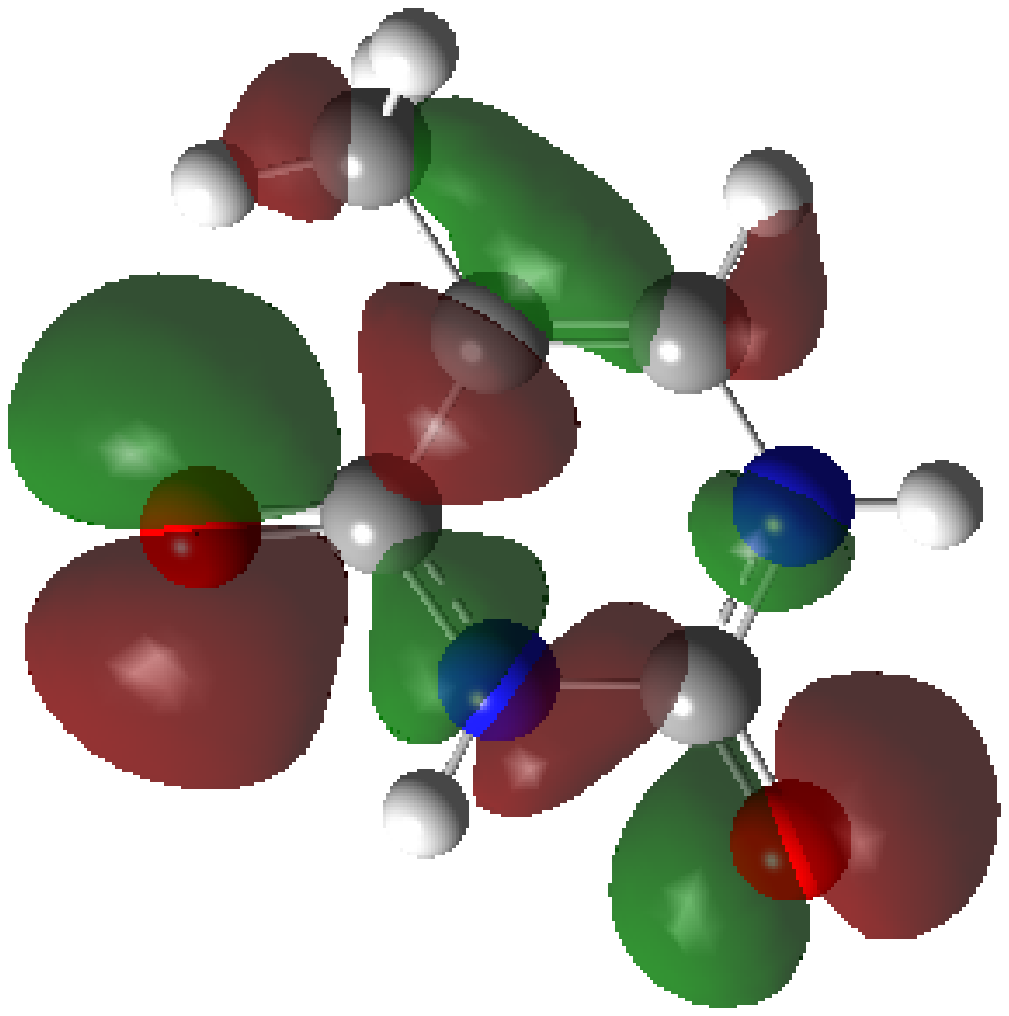}
\includegraphics[angle=0,width=1.1cm]{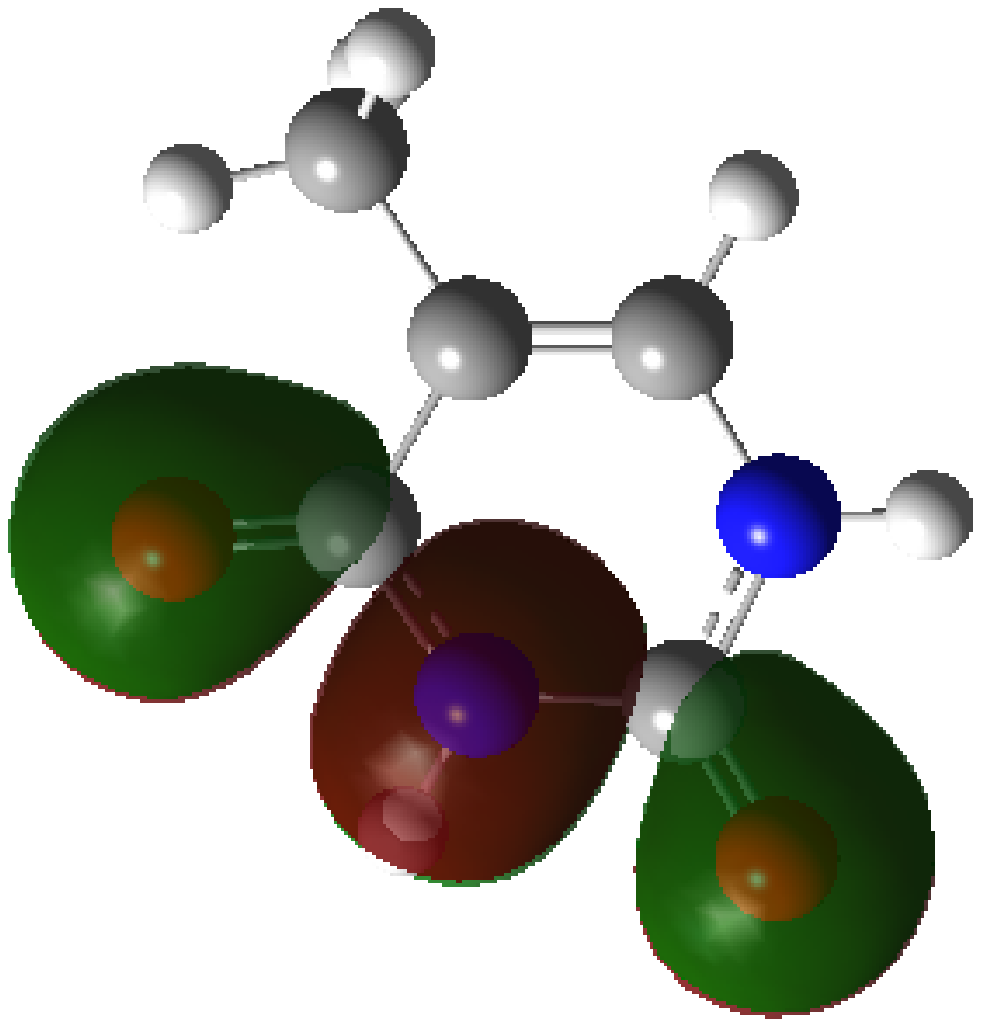}
\includegraphics[angle=0,width=1.1cm]{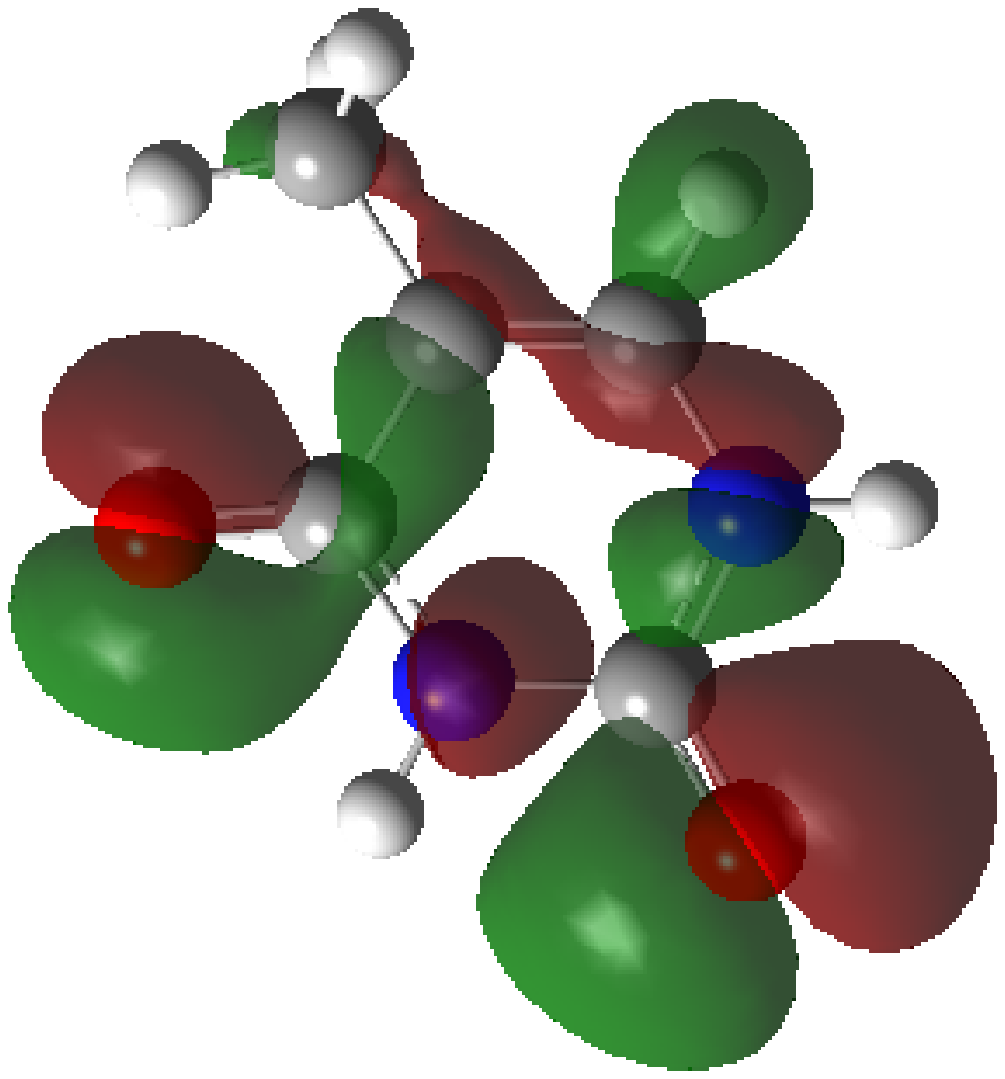}
(b)
\includegraphics[angle=0,width=1.1cm]{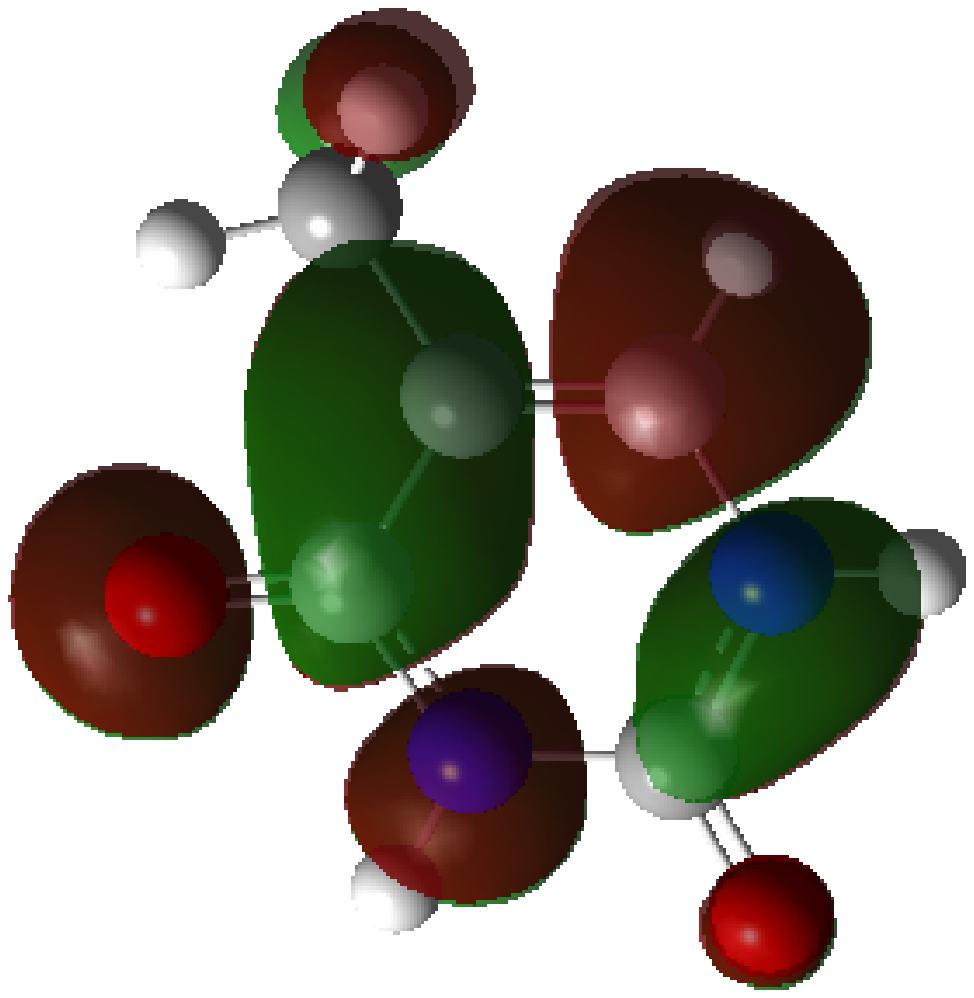}
\includegraphics[angle=0,width=1.1cm]{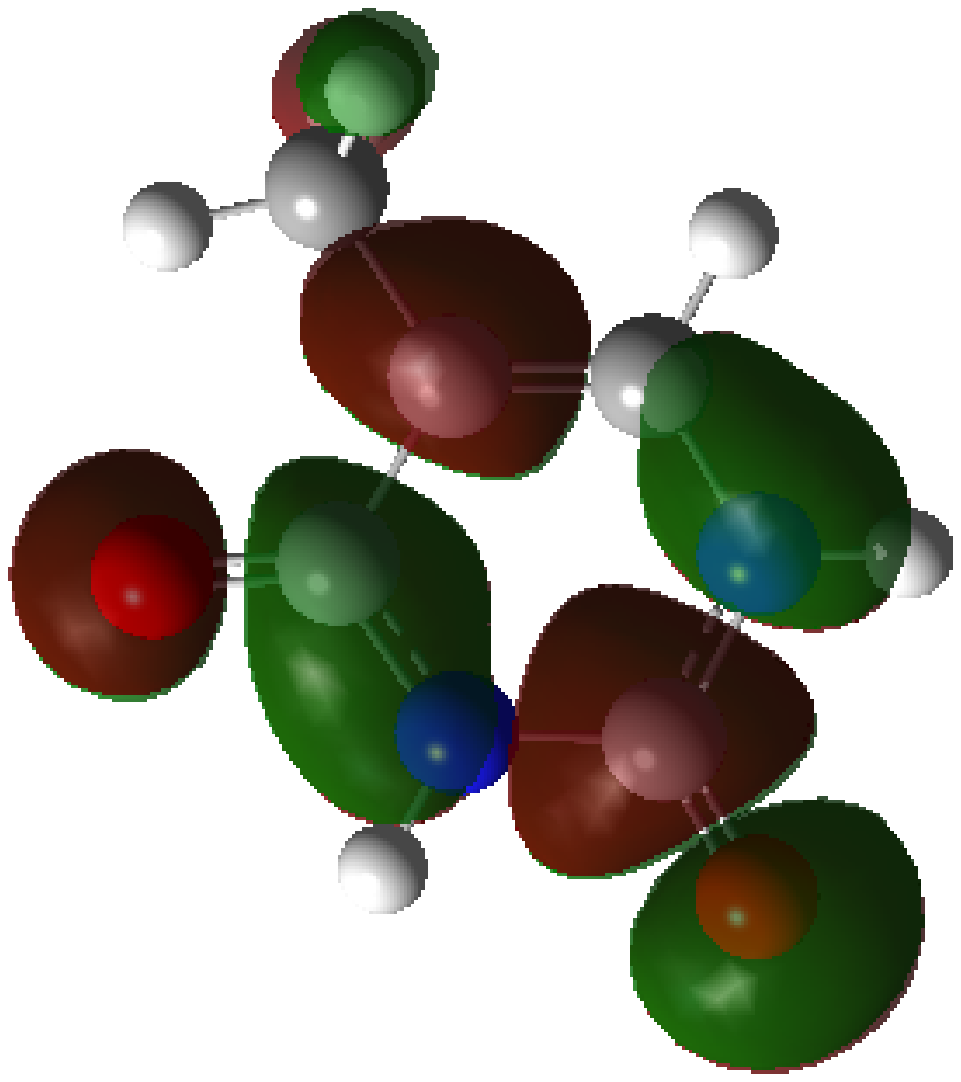}\\
\raggedright
\hspace{8ex}P1\hspace{6ex}N1\hspace{6ex}P2\hspace{6ex}N2\hspace{10ex}S1\hspace{6ex}S2
\caption{Standard-orbitals generated in the Thy B3LYP-DFT calculation.  P1, N1, P2 and N2 correspond to the HOMO, HOMO-1, HOMO-2, and HOMO-3, respectively, and S1 and S2 are LUMO and LUMO+1.  P1 and P2 (S1 and S2) are the major components of the h- (e-)orbital of $^{1}\pi\pi^{*}$-character excitations, while N1 and N2 are those of h-orbital of $^{1}n\pi^{*}$-character excitations.}
\end{figure}

\begin{table}
\caption{
Electronic transition origin (shown in the NTO1 column) of the first 7 SO-hosted singlet excitations
of ideal Thy and dT given by EOM-CCSD and TD-$\omega$B97X calculation.  N denotes the excitation order.  \% records the ratio
of NTO1 to the excitation.  Type denotes the excitation classification based on EOM-CCSD Thy NTO1 expressions;
for EOM-CCSD the total domination of LCSEC (\%) is also listed in the same column.}
\begin{tabular}{|c|clcc|clcc|}
\hline
           &             \multicolumn{ 4}{c|}{Ideal Thy} &              \multicolumn{ 4}{c|}{Ideal dT} \\
\hline
\multicolumn{ 1}{|c|}{ } & \multicolumn{ 1}{c|}{N} & NTO1 & \multicolumn{ 1}{|c|}{\%} &       Type & \multicolumn{ 1}{c|}{N} &  NTO1 & \multicolumn{ 1}{|r|}{\%} &       Type \\

\hline
      EOM- &     1 &    N1 - S1 &   85 &  A(86) &    1 &   N1 - S1 &    79 &   A(80) \\

      CCSD &     2 &    P1 - S1 &   83 &  B(83) &    2 &   P1 - S1  &   83  &  B(83) \\

           &     3 & N1 N2 -   &    86 &  C(88) &    3 & N1 N2 -  &     72 &   C(75) \\
           &      & (\underline{S1}) S2         &     &   &      & (\underline{S1}) S2 &     &   \\

           &     4  &  P2 - S1  &   83 & D(83) &  4  &  P2 - S1 &   69 & D(83) \\

           &     5  &  P1 - S2  &   86 & E(86) &  5  &  P1 - S2 &   75 & E(84) \\

           &     6  &  N2 - S1  &   83 & F(85) &  6 & N2 (B) - S1 &   78 & { (F)(81)} \\

           &     8 & N1 (\underline{N2})&  72 &    G(84) &  7 & N1 (\underline{N2}) - &  71 &    (G)(75) \\
           &            & - S2         &     &   &      & (\underline{S1}) S2 &     &   \\
\hline
       TD- &     1 &  N1 - S1 &   100 &   A &    1 & N1 - S1 (S2) &         99 &        (A) \\

    $\omega$B97X & 2 &  P1 - S1 &    97 &   B &    2 & P1 - S1 &   97 &    B \\

           &     3 & N1 N2 -  &    99 &   C &    3 & N1 N2 - &   98 &    C \\
           &       & (\underline{S1}) S2         &     &   &      & (\underline{S1}) S2 &     &   \\

           &     4 &  P2 - S1 &    99 &   D &    4 & P1 - S2 &   96 &    E \\

           &     5 &  P1 - S2 &    98 &   E &    5 & P2 - S1 &   97 &    D \\

           &    6 &   N2 - S1 &    97 &   F &    6 & N2 B (O) &  96 & (F) \\

           &            &           &     &   &      & - S1 &     &   \\
           &    7 & N1 \underline{N2} -  &   91 &  (G) &   7 & N1 \underline{N2} - &  83 & (G) \\

           &      & (\underline{S1}) S2         &     &   &      & (\underline{S1}) S2 &     &   \\

\hline
\end{tabular}
\end{table}

The results of several chosen excitations of Thy and dT are shown in Fig. 3, Table II and ~\cite{R-Supp}.
The chosen excitation, regarded as SO-hosted (standard-orbitals hosted), has its NTO1-H and NTO1-E overall more than 30\% density
contributions from the monomer orbitals (standard-orbitals) shown in Fig. 2(a) and (b), respectively.
To confirm this, their projection to standard-orbitals is first performed and the density contribution of a standard-orbital is given
by the square of its coefficient; a transition expression is also determined from the coefficients (detailed in ~\cite{R-Supp}).
When dealing with dT, we also consider the involvement of all orbitals originated from the backbone structure
(denoted as 'B') to a NTO1(2)-H.  The integration of standard-orbitals (plus 'B' for dT) generally accounts for
more than 90\% of a NTO1(2)-H(E).  Otherwise, we attribute the involvement of all other Thy-orbitals as 'O'.
Moreover, a precise statistical assessment of NTO1 of an excitation is carried out as well, where the root mean
square deviation of the projection coefficients ($\sigma$) of NTO1-H and NTO1-E of each electronic transition
is computed.  $\sigma_{M}$ is used when we map the NTO1 obtained from TDDFT with various functionals onto NTO1 of
EOM-CCSD, and $\sigma_{E}$ is for comparison between EOM-CCSD results for Thy and dT.

\begin{figure}
\raggedright
(a)\hspace{20ex}(b)\hspace{20ex}(c)
\includegraphics[angle=0,width=2.8cm]{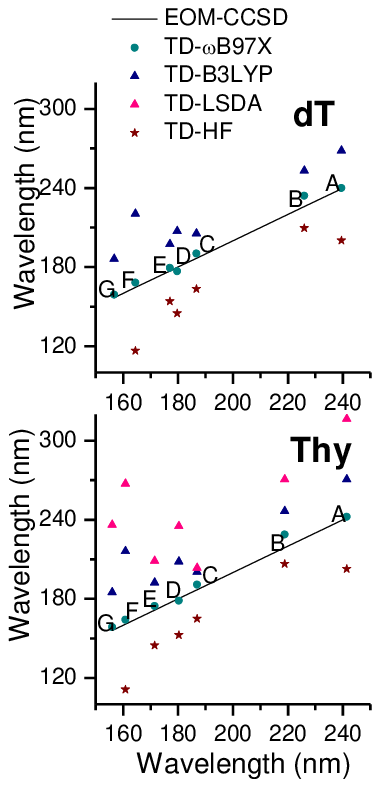}
\includegraphics[angle=0,width=2.8cm]{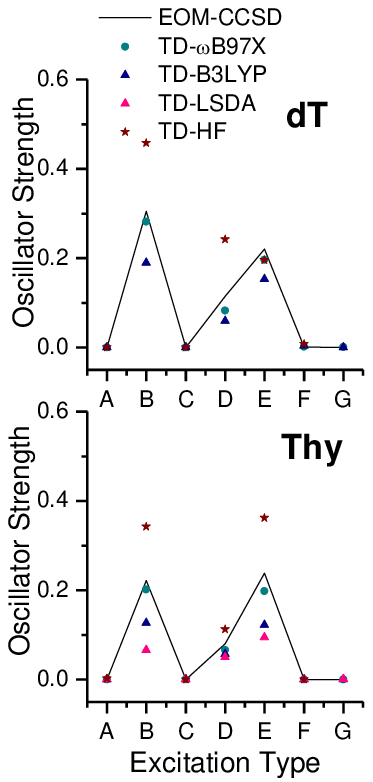}
\includegraphics[angle=0,width=2.8cm]{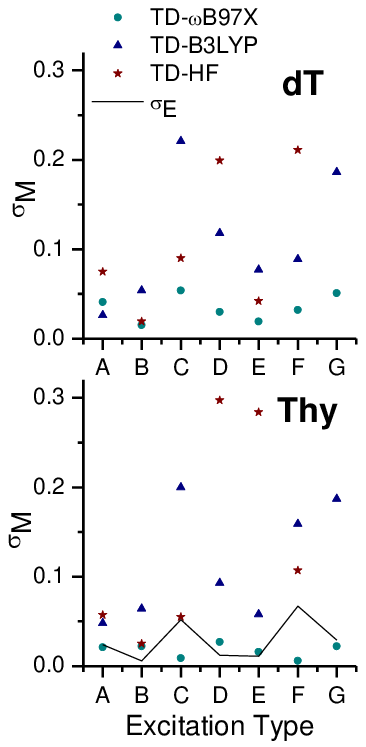}
\caption{
(a) Absorption wavelengths calculated using TD-LDA (pure functional), TD-HF (exact exchange), TD-B3LYP (global hybrid functional) and TD-$\omega$B97X (LC hybrid functional) vs. absorption wavelengths from EOM-CCSD.  Data of each method for the same excitation lies on the same vertical line.  (b) Oscillator Strength: line denotes the results from EOM-CCSD calculation.  (c) $\sigma_{M}$ of the Type-A$\sim$G calculated by the same methods except TD-LDA; $\sigma_{E}$ is also plotted along with $\sigma_{M}$ in the Thy result.  Type-G is not predicted by TD-HF calculation for Thy whereas the excitations of dT given by TD-LDA have quite different transition origin and cannot be well-classified as Type-A$\sim$Type-G~\cite{R-Supp}.}
\end{figure}

From Table II, it can firstly be seen in EOM-CCSD results that the low-lying excitations indeed can be well described by LCSEC,
as its total domination to whole excitation is generally higher than 0.80 (0.75 to be the lowest); secondly their LCSEC
can generally be well explained by NTO1, as its domination is often close to the LCSEC one (NTO2 may arise for higher-lying
excitations or the dimer systems mentioned earlier).  TD-$\omega$B97X (and other LC-TDDFT) results agree quite
well with those of EOM-CCSD, even for the small system Thy, not only in terms of the absorption energy, oscillator strength
but also in terms of the predicted NTO1 transition origin (Fig. 3), and thus have notably outperformed other tested theoretical
methods~\cite{R-Supp}.  Incidentally, one should note that EOM-CCSD provides generally smaller
NTO1 domination than TDDFT does even for the lowest excitation of Thy.  This is attributed to the limitation of adiabatic
approximation of exchange-correlation kernel used in this work~\cite{R-omegaTD}. In other words, TDDFT renders LCSEC to
be 100\% of an excitation.  Furthermore, LCSEC from LC-TDDFT agrees well with that of EOM-CCSD in that
the relative percentage of different SECs is well reproduced. Therefore,
the $\sigma_{M}$ of LC-TDDFT for the low energy absorptions in Table II is very small.

\setlength{\parskip}{0cm}
The key factor causing functionals to have the results of different quality is their HF-exchange prescription,
as shown in ~\cite{R-Supp}.  Global hybrid functionals, like B3LYP and PBE1PBE, bear increased self-interaction
error (SIE) for charge-transfer transition component with systematic underestimation of transition energy.
Instead, LC hybrid functionals, implemented with a distance-dependent HF-exchange, can efficiently reduce SIE.
In ~\cite{R-LC5}, it has been argued that LDA and GGAs (semilocal density functionals) are accurate only for small
interelectronic separations (r).  Even though how precisely the amount of HF-exchange should change with $r$ remains
unknown~\cite{R-LC9}, the importance of HF-exchange generally grows with an increasing $r$.  Compared with the results
of other theoretical methods demonstrated in ~\cite{R-Supp} we find that TD-$\omega$B97X, having the smallest average
$\sigma_{M}$, best reproduces the NTO1 transition origin that EOM-CCSD predicts for
the several low-lying excitations of Thy and dT without sacrificing other excitation properties.
Hence it is a very good choice to use when studying the electronic excitations of DNA systems.  For other molecular
systems more tests may still be needed; for open-shell systems the use of present model also requires extra
consideration where the ground state wavefunction is often no-longer dominated by a single Slater determinant.
\setlength{\parskip}{0cm}

For electronic excitations, QNTO analysis provides a definite assessment of their transition
origin which has direct connection to the electronic structure upon which later molecular dynamics
(nuclear motions, non-radiative transitions, photo-induced chemical reactions, and etc.) depend.
When one deals with several electronic excitations of a molecule with low or even no symmetry, it also provides the
fingerprint for them.
A measure of transition origin variation of particular electronic excitations
(e.g. SO-hosted ones in this work) due to different implemented theoretical methods ($\sigma_{M}$) or
different environments ($\sigma_{E}$) can be performed as well.  Since the computational cost is moderate,
LC-TDDFT cooperated with QNTO analysis is expected to be particularly useful in studying larger systems.

We thank National Science Council and NCTS of Taiwan for supports.


\begin{table*}[t]
\large{\textbf{Supplemental Material: The Quantified NTO Analysis for the Electronic Excitations of Molecular Many-Body Systems}}
\end{table*}

\makeatletter
\setcounter{figure}{0}
\setcounter{table}{0}
\renewcommand\thefigure{S\@arabic\c@figure}
\renewcommand\thetable{S\@Roman\c@table}
\renewcommand{\bibnumfmt}[1]{[S#1]}
\renewcommand{\citenumfont}[1]{S#1}

\begin{figure}[H]
\raggedright
\hspace{6ex}u\hspace{18ex}v\\
\includegraphics[angle=0,width=4.5cm]{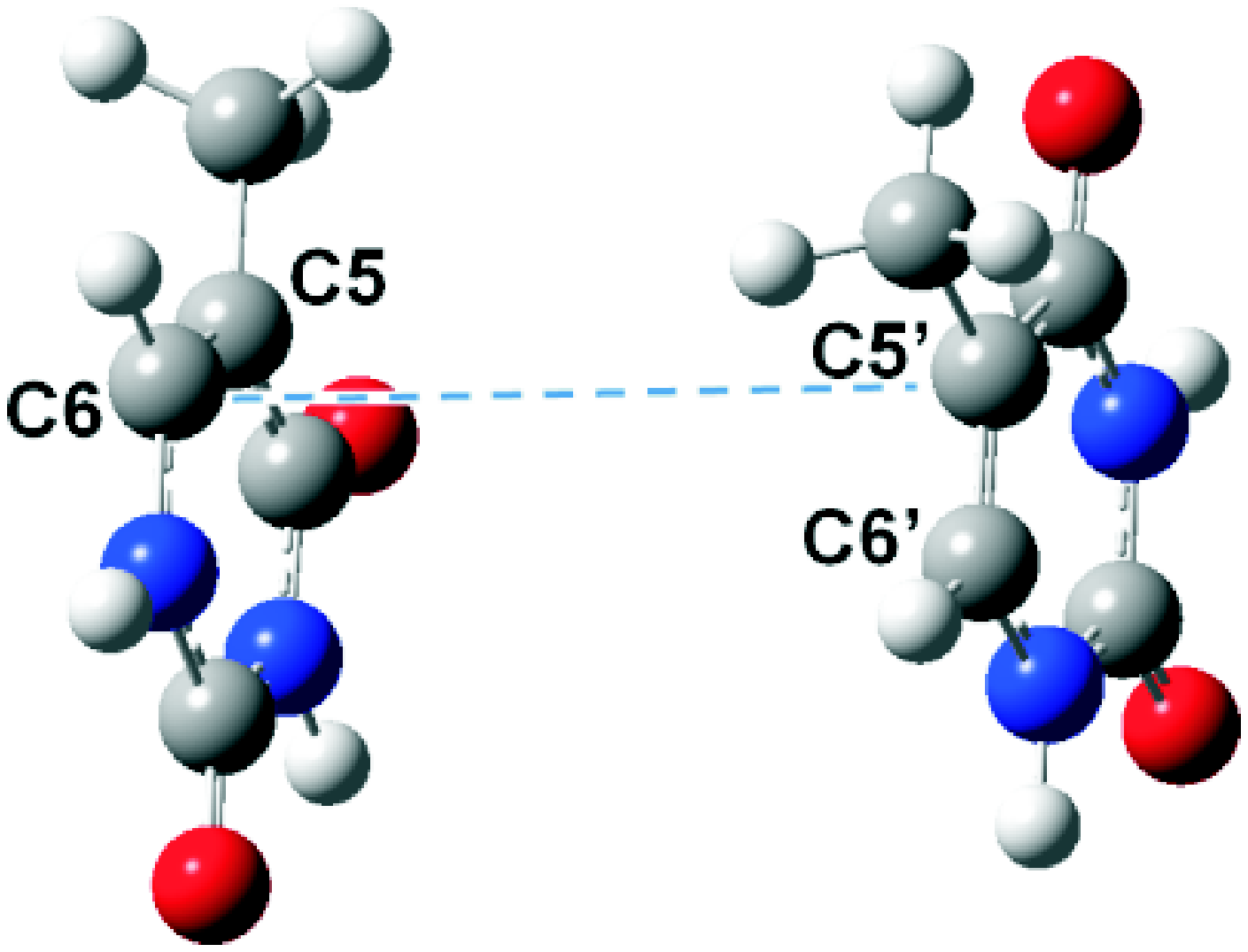}
\caption{
The Thy dimer used in Table I for the demonstration of molecular exciton theory recovery by quantified natural transition orbitals (QNTO) analysis.  In the process the distance of C6-C5' (dash line) is modified with the two Thy base planes bearing unchanged polar angle and azimuthal angle in spherical coordinate system of z-axis along C6-C5'.  In other words, the transition-dipole-transition-dipole-product of an excitation, a quantity on which the exciton theory depends, is fixed and only the distance between the two transition-dipoles is differentiated.
}
\end{figure}

\begin{figure}[H]
\raggedright
\includegraphics[angle=0,width=8cm]{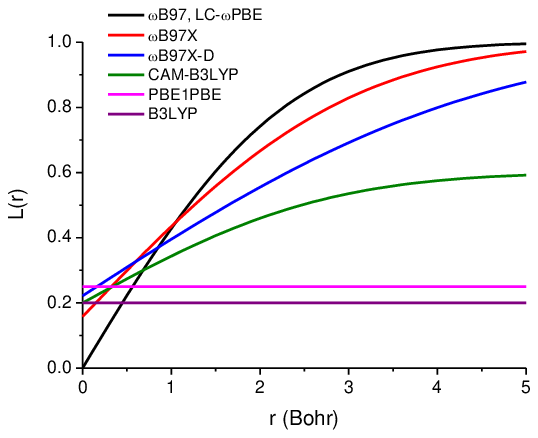}
\caption{
L(r) vs. r for various functionals.  L(r), defined as (HF operator * r), is a fraction of Hartree-Fock exchange at interelectronic separation r.
}
\end{figure}

\begin{figure}[H]
\raggedright
(a)

\includegraphics[angle=0,width=4cm]{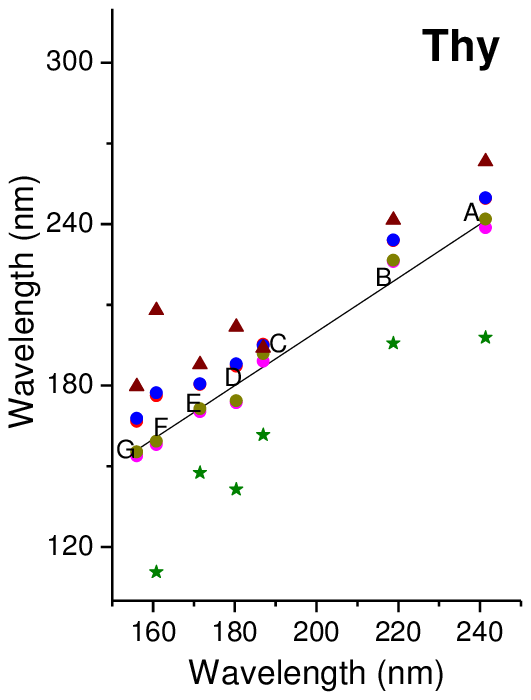}
\includegraphics[angle=0,width=4cm]{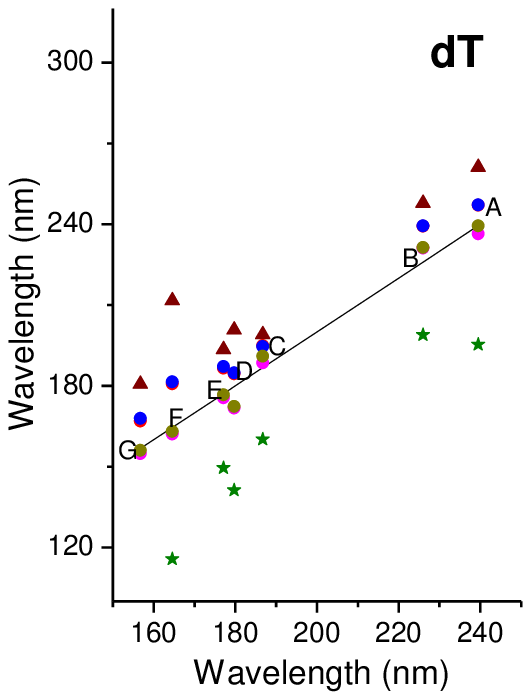}

(b)

\includegraphics[angle=0,width=4cm]{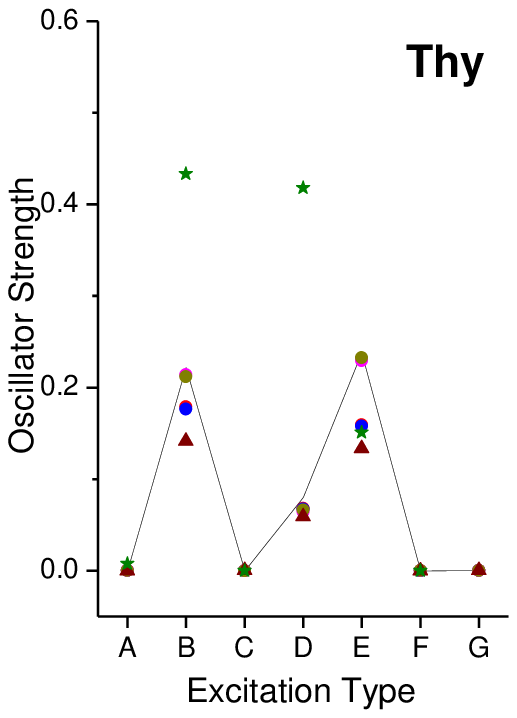}
\includegraphics[angle=0,width=4cm]{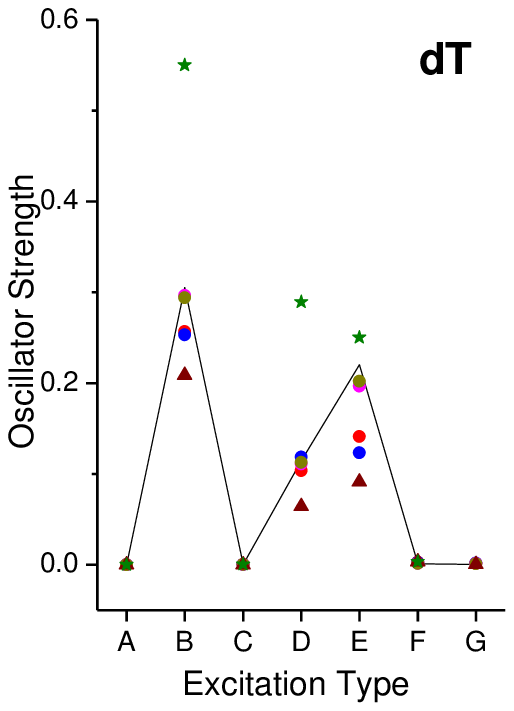}

(c)

\includegraphics[angle=0,width=4cm]{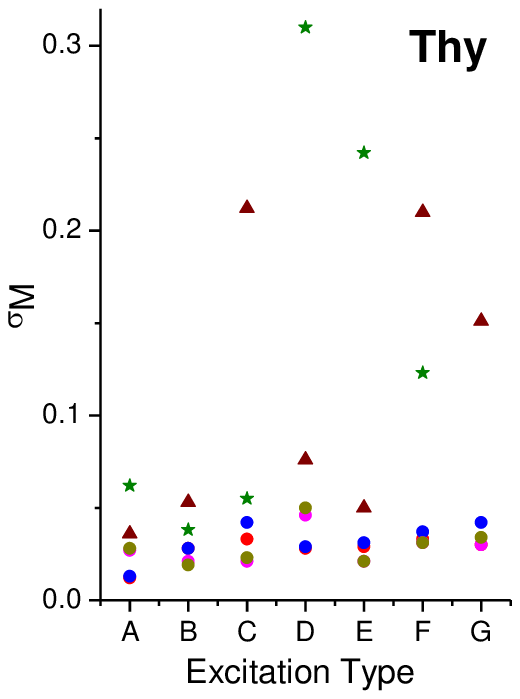}
\includegraphics[angle=0,width=4cm]{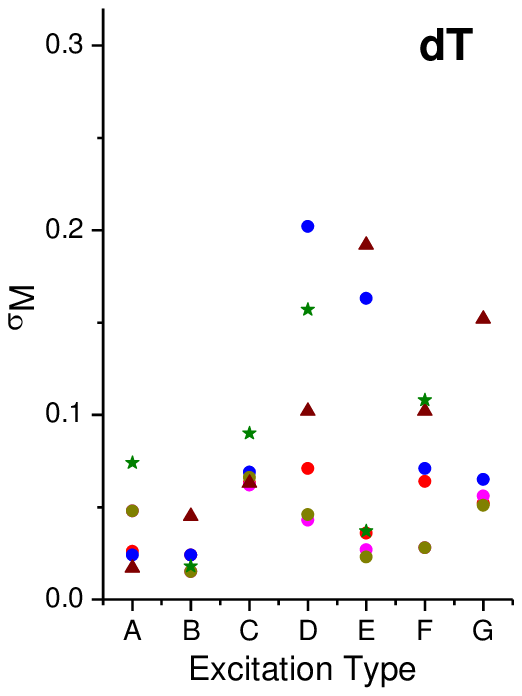}
\caption{
The results of more theoretical methods not mentioned in Fig. 3.  Notations are detailed in the main text. (a)Absorption wavelengths calculated using TD-CAM-B3LYP({\color{red}$\bullet$})~\cite{R-CAM}, TD-$\omega$B97X-D({\color{blue}$\bullet$})~\cite{R-wB97XD}, TD-$\omega$B97({\color{magenta}$\bullet$})~\cite{R-wB97}, TD-LC-$\omega$PBE({\color{darkyellow}$\bullet$})~\cite{R-LCwPBE} (LC hybrid functionals), TD-PBE1PBE({\color{wine}$\blacktriangle$})~\cite{R-PBE1PBE} (a global hybrid functional) and CIS({\color{olive}$\star$})~\cite{R-CIS} (TD-HF with Tamm-Dancoff approximation) vs. absorption wavelengths from EOM-CCSD($\diagup$)~\cite{R-EOM}.  Data of each method for the same excitation lies on the same vertical line.  (b) Oscillator Strength: line denotes the results from EOM-CCSD calculation.  (c) $\sigma_{M}$ of the Type-A$\sim$Type-G excitations.
}
\end{figure}

\setcounter{page}{1}
\renewcommand\thepage{S\arabic{page}}

\begin{table}[H]
\caption{{
Electronic transition origin of the first 8 SO-hosted singlet excitations of ideal Thy and dT given by calculation
of (a) TDDFT with four different LC hybrid functionals: CAM-B3LYP, $\omega$B97, $\omega$B97X-D and LC-$\omega$PBE and
(b) other theoretical methods: CIS, TD-HF, TD-LDA~\cite{R-LDA1,R-LDA2} and TD-PBE1PBE not demonstrated in
Table II.  Incidentally, in CIS result of dT only 7 SO-hosted excitations out of the first 50 states
(to 99.14nm for the 50th one) are recognized and listed.  N and P, respectively, denote the excitation order and
the phase of NTO2 for the absorptions with NTO1 domination to the linear combination of singly excited configuration (LCSEC)
lower than 70\%.  The next column shows the transition origin of NTO1(2), followed by the NTO1(2) domination to the
excitation.  The last column denotes the excitation classification based on EOM-CCSD Thy NTO1 expressions.
The excitation denoted as (F') has similar Type-F NTO1 expression
but larger $\sigma_{M}$.}}
{\fontsize{6pt}{12pt}\selectfont
\raggedright
\begin{tabular}{|c|clcc|clcc|}
\hline
       (a)    &             \multicolumn{ 4}{c|}{Ideal Thy} &              \multicolumn{ 4}{c|}{Ideal dT} \\
\hline

\multicolumn{ 1}{|c|}{} & \multicolumn{ 1}{c|}{N;} & NTO1 & \multicolumn{ 1}{|c}{\%} & \multicolumn{ 1}{|c|}{Type} & \multicolumn{ 1}{c|}{N;} & NTO1(2) & \multicolumn{ 1}{|c}{\%} & \multicolumn{ 1}{|r|}{Type} \\

\multicolumn{ 1}{|c|}{} & \multicolumn{ 1}{c|}{P} &   & \multicolumn{ 1}{|c}{} & \multicolumn{ 1}{|c|}{} & \multicolumn{ 1}{c|}{P} &   & \multicolumn{ 1}{|c}{} & \multicolumn{ 1}{|c|}{} \\
\hline
       TD- &          1 &    N1 - S1 &        100 &          A &          1 &    N1 - S1 &         99 &          A \\

      CAM- &          2 &    P1 - S1 &         97 &          B &          2 &    P1 - S1 &         97 &          B \\

     B3LYP &          3 & N1 N2 - (\underline{S1}) S2 &         98 &          C &          3 & N1 N2 - (\underline{S1}) S2 &         97 &          C \\

           &          4 &    P2 - S1 &         95 &          D &          4 &    P1 - S2 &         69 &          E \\

           &          5 &    P1 - S2 &         94 &          E &          - &    P2 - S1 &         29 &            \\

           &          6 & N2 - S1 (S2) &       95 &        (F) &          5 &    P2 - S1 &         71 &          D \\

           &          7 & N1 \underline{N2} - S2 &         94 &        (G) &          6 & N2 B - S1 (S2) &         92 &        (F) \\

           &          9 & P2 - S2 &   93 &    - &          7 &  N2 B - S1 &         91 &       (F') \\

           &            &            &            &            &          8 & N1 \underline{N2} - S2 &         86 &        (G) \\
\hline
       TD- &          1 &    N1 - S1 &        100 &          A &          1 &    N1 - S1 &         99 &          A \\

    $\omega$B97X &     2 &   P1 - S1 &         96 &          B &          2 &    P1 - S1 &         97 &          B \\

     -D  &          3 & N1 N2 - (\underline{S1}) S2 &         97 &          C &          3 & N1 N2 - (\underline{S1}) S2 &         96 &          C \\

        &       4 &    P2 - S1 &       94 &       D &        4 &     P1 (P2) - (\underline{S1}) S2 &       56 &  (E) \\

           &       5 &    P1 - S2 &       93 &        E &      - &   (\underline{P1}) P2 - S1 (S2) &       43 &    \\

           &          6 & N2 - S1 (S2) &         94 &        (F) &        5 &    (P1) P2 - S1 (S2) &       59 &   (D) \\

           &          7 & N1 \underline{N2} - S2 &         94 &        (G) &    + &    P1 (\underline{P2}) - (\underline{S1}) S2 &     39  &        \\

           &          9 & P2 - S2 &   94 &     - &          6 & N2 B - S1 (S2) &         91 &        (F) \\

           &            &            &            &            &          7 &  N2 B - S1 &         89 &       (F') \\

           &            &            &            &            &          8 & N1 \underline{N2} - S2 &         83 &        (G) \\
\hline
       TD- &          1 & N1 - S1 (S2)  &         99 &        (A) &          1 & N1 - S1 (S2) &         99 &        (A) \\

      $\omega$B97 &       2 &   P1 - S1 &       97 &         B &         2 &   P1 - S1 &      97 &       B \\

      &          3 & N1 N2 - S2 &         99 &        (C) &          3 & N1 N2 - S2 &         98 &        (C) \\

      &      4 &   P2 - S1 &     94 &      D &      4 &    P1 - S2 &     90 &      E \\

           &      5 &   P1 - S2 &     93 &      E &      5 &   P2 - S1 &     91 &      D \\

           &          6 & N2 (O) - S1 &         94 &        (F) &          6 & N2 (B) (O) - S1 &         96 &        (F) \\

           &          7 & N1 (\underline{N2}) - (\underline{S1}) S2 &         85 &        (G) &          7 & N1 (\underline{N2}) - (\underline{S1}) S2 &         81 &        (G) \\

           &          9 & P2 - S2 &   94 &    - &          8 & (N2) B (O) - S1 &         80 &       (F') \\
\hline
    TD-LC- &          1 & N1 - S1 (S2) &        100 &        (A) &          1 & N1 - S1 (S2) &         99 &        (A) \\

      $\omega$PBE &      2  &   P1 - S1  &     97  &      B  &      2  &   P1 - S1  &     97  &      B  \\

      &          3 & (N1) N2 - S2 &         99 &        (C) &          3 & (N1) N2 - S2 &         99 &        (C) \\

      &      4  &   P2 - S1  &     94  &      D  &      4  &   P1 - S2  &     92  &      E  \\

           &      5  &   P1 - S2  &     92  &      E  &      5  &   P2 - S1  &     93  &      D  \\

           &          6 & N2 (O) - S1 &         95 &        (F) &          6 & N2 (B) (O) - S1 &         97 &        (F) \\

           &          7 & N1 (\underline{N2}) - (\underline{S1}) S2 &         85 &        (G) &          7 & N1 (\underline{N2}) - (\underline{S1}) S2 &         83 &        (G) \\

           &          8 & P2 - S2 &   94 &    - &          8 & (N2) B (O) - S1 &         78 &       (F') \\
\hline
\end{tabular}
}
\end{table}

\begin{table*}
{\fontsize{6pt}{12pt}\selectfont
\begin{tabular}{|c|clcc|clcc|}
\hline
        (b)   &             \multicolumn{ 4}{c|}{Ideal Thy} &              \multicolumn{ 4}{c|}{Ideal dT} \\
\hline
\multicolumn{ 1}{|c|}{} & \multicolumn{ 1}{c|}{N;} & NTO1(2) & \multicolumn{ 1}{|c}{\%} & \multicolumn{ 1}{|c|}{Type} & \multicolumn{ 1}{c|}{N;} & NTO1(2) & \multicolumn{ 1}{|c}{\%} & \multicolumn{ 1}{|r|}{Type} \\

\multicolumn{ 1}{|c|}{} & \multicolumn{ 1}{c|}{P} &    & \multicolumn{ 1}{|c}{} & \multicolumn{ 1}{|c|}{} & \multicolumn{ 1}{c|}{P} &    & \multicolumn{ 1}{|c}{} & \multicolumn{ 1}{|c|}{} \\
\hline
       CIS &          1 & N1 (N2) - S1 (S2) (O) &         98 &        (A) &      1  &   P1 - S1  &     94  &    B  \\

           &      2  &   P1 - S1  &     94  &    B &          2 & N1 (N2) - S1 (S2) (O) &         98 &        (A) \\

           &          3 & (N1) N2 - S2 (O) &         99 &        (C) &          3 & (N1) N2 - S2 (O) &         97 &        (C) \\

           &      4  &   P1 P2 - S1 S2  &     59  &    (E)  &      4  &   P1 - S2  &     76  &      E  \\

           &      +  &   P1 P2 - S1 S2  &     35  &       &      5  &   P2 - S1 (S2)  &     81  &    (D)  \\

           &      6  &   P1 P2 - S1 S2  &     73  &    (D)  &    7 & P2 - (S1) S2 &   87 &    - \\

           &         8 & P2 - (S1) S2 &   88 &     - &         15 & (N1) N2 O - S1 &         45 &        (F) \\

           &         12 & N1 N2 (O) - S1 &         64 &       (F') &          - & P1 (B) - O &         26 &            \\

           &          - & N1 N2 (O) - S2 (O) &         32 &            &            &            &            &            \\

           &         15 &  N2 O - S1 &         68 &        (F) &            &            &            &            \\

           &          + & N1 (N2) O - S2 (O) &         23 &            &            &            &            &            \\
\hline
     TD-HF &      1  &   P1 - S1  &     97  &      B  &      1  &   P1 - S1  &     97  &      B  \\

           &          2 & N1 (N2) - S1 (S2) (O) &         99 &        (A) &          2 & N1 (N2) - S1 (S2) (O) &         98 &        (A) \\

           &          3 & (N1) N2 - S2 (O) &        100 &        (C) &          3 & (N1) N2 - S2 (O) &         98 &        (C) \\

           &      4  &   P1 P2 - S1 S2  &     60  &    (D)  &      4  &   P1 - S2  &     71  &      E  \\

           &      +  &   P1 P2 - S1 S2  &     36  &       &      5  &   P2 - S1 S2  &     80  &    (D)  \\

           &      5  &   P1 P2 - (S1) S2  &     77  &    (E)  &   7 & P2 - (S1) S2 &   88 &    - \\

           &    8 & P2 - (S1) S2 &   88 &    - &         15 & (P1) (N2) (B) (O) - S1 O &         38 &        (F) \\

           &         15 &  N2 O - S1 &         71 &        (F) &          + & P1 (N1) (N2) (O) - S1 O &         32 &            \\

           &   22 & P1 O - S2 O &   39 &    - &         21 & P1 - S1 (O) &         82 &          - \\

           &   + & N1 (N2) - O &   32 &      &            &            &            &            \\
\hline
       TD- &          1 &    N1 - S1 &        100 &          A &          1 &    N1 - S1 &        100 &          A \\

     B3LYP &      2  &   P1 - S1  &     96  &      B  &      2  &   P1 - S1  &     97  &      B  \\

           &          3 &    N2 - S1 &         93 &          F &          3 &  N2 B - S1 &         96 &        (F) \\

           &      4  &   P2 - S1  &     92  &      D  &      4  &   P2 - S1  &     81  &      D  \\

           &          5 & N1 (N2) - S2 &         86 &        (C) &          5 & (N1) (N2) B - S1 (S2) &         63 &        (C) \\

           &      6  &   P1 - S2  &     89  &      E  &      +  & N1 B - (S1) S2 &         36 &       \\

           &          7 & (N1) N2 - S2 &         96 &        (G) &          6 & N1 N2 - S1 S2 &         61 &       (CG) \\

           &          9 & P2 - S2 &   82 &    - &          + & N1 (N2) B - S1 S2 &         39 &            \\

           &            &            &            &            &      7  &   P1 - S2  &     75  &      E  \\

           &            &            &            &            &          9 & (N1) N2 - S2 &         96 &        (G) \\
\hline
       TD- &          1 &    N1 - S1 &        100 &          A &          1 &    N1 - S1 &        100 &          A \\

     PBE1- &      2  &   P1 - S1  &     96  &      B  &      2  &   P1 - S1  &     97  &      B  \\

       PBE &          3 & (N1) N2 - S1 (S2) &         90 &        (F) &          3 &  N2 B - S1 &         94 &        (F) \\

           &      4  &   P2 - S1  &     93  &      D  &      4  &   P2 - S1  &     80  &      D  \\

           &          5 & N1 N2 - (S1) S2  &         80 &        (C) &          5 & N1 (N2) - (S1) S2 &         61 &        (C) \\

           &      6  &   P1 - S2  &     90  &      E  &          - & B - S1 (S2) &         37 &            \\

           &          7 & (N1) N2 - S2 &         95 &        (G) &      6  &   P1 (N1) (N2) - (S1) S2  &     69  &  (E)  \\

           &          9 & P2 - S2 &   90 &     - &      +  &   (P1) (N2) P2 (B) - S1 (S2)  &     30  &       \\

           &            &            &            &            &          7 & (P1) N2 (B) - S1 (S2) &         64 &        (F) \\

           &            &            &            &            &          - & P1 (N1) B - (S1) S2 &         35 &            \\

           &            &            &            &            &          9 & (N1) N2 - S2 &         92 &        (G) \\
\hline
       TD- &          1 &    N1 - S1 &        100 &          A &          1 &    N1 - S1 &        100 &          A \\

      LDA &      2  &   P1 (P2) - S1  &     97  &    (B)  &      2  &   (P1) B - S1  &     99  &    (B)  \\

           &          3 &    N2 - S1 &        100 &          F &      3  &   P1 N2 (B) - S1  &     98  &   (BF)  \\

           &          4 &    N1 - S2 &         98 &        (G) &          5 & (P1) N2 B - S1 &         99 &        (F) \\

           &      5  &   (P1) P2 - S1 (S2)  &     92  &   (DB)  &          6 &    N1 - S2 &         99 &        (G) \\

           &      6  &   P1 - S2  &     85  &      E  &      7  &   (P1) P2 - S1 (S2)  &     85  &   (DB)  \\

           &          7 &    N2 - S2 &        100 &        (C) &      9  &   (P1) (N2) B - S2  &     86  &    (E)  \\

           &         11 &    P2 - S2 &         48 &      - &     10  &   P1 (N2) (B) - S2 (O)  &     81  &    (E)  \\

           &         + &    (P1) O - S1 &      44 &        &       &       &       &       \\
\hline
\end{tabular}
}
\end{table*}
\textbf{Note for Table SI:}

The coefficient threshold to output singlet excited configurations (SECs) for QNTO analysis in TDDFT calculations is set to 0.01; for EOM-CCSD it is set to 0.1.  Even though the threshold used is different, we assume the SECs with coefficient smaller than 0.1 have a minor effect on the composition of LCSEC so that EOM-CCSD results can still be used as a norm for TDDFT results for comparison.  In Table SI NTO1(2) is the acronym of the (2nd) dominant natural transition orbital pair.  P1, N1 and etc. are the standard-orbitals defined in Fig. 2 used to expand the hole- (NTO1(2)-H) and electron-orbital (NTO1(2)-E) of a NTO1(2).
To determine the transition expression of a NTO1(2), P1, for example, is involved, partial-involved and not involved in a NTO1-H if the $|\langle P1|NTO1-H\rangle|^{2}$, density contribution of P1 to the NTO1-H, is $[1, 0.3]$, $(0.3, 0.1]$ and $(0.1, 0]$ respectively; $\langle P1|NTO1-H\rangle$ denotes the projection coefficient of the NTO1-H onto P1.  Based on these coefficients the SO-hosted excitations, assigned as having NTO1-H and NTO1-E overall more than 30\% density contributions from the standard-orbitals shown in Fig. 2(a) and Fig. 2(b), respectively, can also be readily identified from several excitations calculated by a theoretical method.  Inside "()" in a transition expression indicates the standard-orbital that has partial-involvement; the underline below a standard-orbital means its coefficient is negative.  Moreover, when dealing with dT we also introduce 'B' referring to all orbital-fraction of NTO1-H from the DNA sugar backbone attaching to Thy.  If the integrated density contribution of standard-orbitals (plus 'B' for dT) is less than 90\% to a NTO1(2)-H(E), the remaining part that is from other Thy-orbitals is denoted as 'O' in the transition expression (shown in "()" if its contribution is $(0.3, 0.1]$).

It also requires extra cautions if the comparison of two transition expressions with some phase difference of standard-orbitals is being carried out. For instance, paying attention on the TD-PBE1PBE result, one finds that the NTO1 represented by "N1 N2 - (S1) S2" for the 5th excitation of Thy is not fully equal to the NTO1, "N1 (N2) - (\underline{S1}) S2", for the 5th excitation of dT because of the S1 phase difference; even single-particle operator, e.g. $\hat{\rho}(\mathbf{r})$, can recognize the difference.  It can be proved that $\langle\Psi^{a}_{i}+\Psi^{b}_{i}|\hat{\rho}(\mathbf{r})|\Psi^{a}_{i}+\Psi^{b}_{i}\rangle\neq\langle\Psi^{a}_{i}-\Psi^{b}_{i}|\hat{\rho}(\mathbf{r})|\Psi^{a}_{i}-\Psi^{b}_{i}\rangle$ because $\langle\Psi^{a}_{i}|\hat{a}^{\dagger}_{a}\hat{a}_{b}|\Psi^{b}_{i}\rangle$ connects the cross terms.  On the other hand, the relative phase between NTO1 and NTO2 can be absorbed into NTO1 or NTO2.  For example, the complete expression for the 7th excitation of dT in the TD-PBE1PBE result reads $\sqrt{0.64}$"(\underline{P1}) N2 (B) - S1 (S2)"-$\sqrt{0.35}$"P1 (\underline{N1}) B - (\underline{S1}) S2".  Absorbing the relative phase between NTO1 and NTO2 into NTO2-H, for example, it can be rewritten as $\sqrt{0.64}$"(\underline{P1}) N2 (B) - S1 (S2)"+$\sqrt{0.35}$"\underline{P1} (N1) B - (S1) \underline{S2}".  ('B' phase is not taken into account.)

\textbf{Results and Discussion}

Among the LC hybrid functionals used in Table SI(a), LC-$\omega$PBE and $\omega$B97 have the same HF-exchange operator (Fig. S2),
while the others have different ones.  LC-$\omega$PBE and $\omega$B97 give quite similar results for both Thy and dT,
showing that the use of different versions of GGA in their functional forms matters little here and the fraction of
HF-exchange with interelectron distance is much more important.  We also see that CAM-B3LYP and $\omega$B97X-D give quite similar results,
which has also been found in ~\cite{R-wB97test} for other molecular systems.

Fig. S3 and Table SI show that TDDFT with LC hybrid functionals evidently outperform other methods for excitation energy, oscillator strength
and NTO1 transition origin prediction.  In general, CIS and TD-HF severely overestimate the excitation energy and oscillator strength,
while TD-B3LYP, TD-PBE1PBE and TD-LDA underestimate them.  The trend that pure density functional (LDA) and global hybrid functionals
(B3LYP, PBE1PBE) tend to give a larger underestimation of oscillator strength for bright absorptions is also observed elsewhere of ~\cite{R-EOMstand}.
The values of $\sigma_{M}$ for different type of excitations yielded by the five methods are in most cases larger than those given by TDDFT
with LC hybrid functionals.  Moreover, LDA, having no HF-exchange in its functional form, leads to excitations with qualitatively
different transition origins that generally cannot be well classified by Type-A$\sim$Type-G.  In fact if we focus on Type-B, D and E (bright absorptions),
their relative absorption energy, oscillator strength, and NTO1 transition origin can be qualitatively captured by TD-B3LYP for both systems.
However, for Type-C, F and G dark absorptions the description of NTO1 transition origin is more problematic and for Type-F its excitation energy
is underestimated more significantly than other absorptions such that its order goes down; both properties play a significant role in later
dynamics of an electronically excited molecule.  TDDFT with LC hybrid functionals, on the other hand, can make them (and also excitation energy,
oscillator strength) agree well with those predicted by EOM-CCSD.

We have observed a close relation between a proper description of electronic excitations and the prescription of HF-exchange in different functionals.  LC hybrid functionals (LC-$\omega$PBE, $\omega$B97, $\omega$B97X, $\omega$B97X-D, CAM-B3LYP) are found to outperform global hybrid functionals (B3LYP, PBE1PBE), pure density functional (LDA) and the exact-exchange (HF) method for excitation energy, oscillator strength and transition origin prediction, when compared with the EOM-CCSD results.  It is also notable that the problematic description of Type-C, F and G excitations by B3LYP and PBE1PBE all concerns with N2-orbital involvement, which is HOMO-3, the deepest used standard-orbital from B3LYP KS-orbitals of Thy, and Type-C, G also concerns with S2-orbital, the LUMO+1, signifying that as (often higher-lying) electronic excitations are concerned with orbitals more far away from HOMO or LUMO, the correction of self-interaction error may be more important even if the system is as small as Thy.

\textbf{Complete reference of Gaussian09}\\
M. J. Frisch, G. W. Trucks, H. B. Schlegel, G. E. Scuseria, M. A. Robb, J. R. Cheeseman, G. Scalmani, V. Barone, B. Mennucci, G. A. Petersson, H. Nakatsuji, M. Caricato, X. Li, H. P. Hratchian, A. F. Izmaylov, J. Bloino, G. Zheng, J. L. Sonnenberg, M. Hada, M. Ehara, K. Toyota, R. Fukuda, J. Hasegawa, M. Ishida, T. Nakajima, Y. Honda, O. Kitao, H. Nakai, T. Vreven, J. A. Montgomery, Jr., J. E. Peralta, F. Ogliaro, M. Bearpark, J. J. Heyd, E. Brothers, K. N. Kudin, V. N. Staroverov, R. Kobayashi, J. Normand, K. Raghavachari, A. Rendell, J. C. Burant, S. S. Iyengar, J. Tomasi, M. Cossi, N. Rega, J. M. Millam, M. Klene, J. E. Knox, J. B. Cross, V. Bakken, C. Adamo, J. Jaramillo, R. Gomperts, R. E. Stratmann, O. Yazyev, A. J. Austin, R. Cammi, C. Pomelli, J. W. Ochterski, R. L. Martin, K. Morokuma, V. G. Zakrzewski, G. A. Voth, P. Salvador, J. J. Dannenberg, S. Dapprich, A. D. Daniels, O. Farkas, J. B. Foresman, J. V. Ortiz, J. Cioslowski, and D. J. Fox, Gaussian 09, Revision A.1, Gaussian, Inc., Wallingford CT, 2009.


\begin{table*}[t]
\caption{{
The standard-orbitals projection coefficients (coefficient square for orbital-fraction from backbone) of NTO1(2)-H(E) of several SO-hosted excitations calculated by (a) EOM-CCSD and TDDFT with LC hybrid functionals, and (b) several other theoretical methods shown in Table II and Table SI.  $\lambda$(nm) and f column record absorption wavelengths and oscillator strengths, respectively.  $\sigma$ column records $\sigma_{E}$ for EOM-CCSD and $\sigma_{M}$ for other theoretical methods.  In addition, there is data of the average $\sigma$ shown in the method column.}}
\raggedright

{\fontsize{6pt}{12pt}\selectfont
\begin{tabular}{|c|ccc|cccc|cc|c||ccc|ccccc|cc|c|}
\hline
     (a) &  \multicolumn{ 3}{c}{Thy} &                   \multicolumn{ 4}{c}{NTO1(2)-H} & \multicolumn{ 2}{c}{NTO1(2)-E} &           &  \multicolumn{ 3}{c}{dT} &                                \multicolumn{ 5}{c}{NTO1(2)-H} & \multicolumn{ 2}{c}{NTO1(2)-E} &           \\

  &          N &      $\lambda$(nm) &       f &         P1 &         N1 &         P2 &         N2 &         S1 &         S2 &   $\sigma$         &          N &      $\lambda$(nm) &       f &         P1 &         N1 &         P2 &         N2 &       (B)$^{2}$ &         S1 &         S2 &   $\sigma$         \\
\hline
      EOM- &          1 &    241.40  &    0.0001  &     -0.01  &      0.96  &     -0.01  &     -0.23  &      0.94  &      0.28  &          - &          1 &    239.47  &    0.0000  &     -0.02  &      0.96  &     -0.03  &     -0.22  &      0.01  &      0.95  &      0.20  &     0.024  \\

      CCSD &          2 &    218.80  &    0.2220  &      0.99  &      0.01  &      0.01  &      0.00  &      0.99  &     -0.01  &          - &          2 &    225.95  &    0.3051  &      0.98  &      0.02  &      0.02  &      0.00  &      0.02  &      0.98  &      0.00  &     0.006  \\

      ($\sigma_{E}$: &          3 &    186.94  &    0.0000  &     -0.01  &      0.59  &     -0.01  &      0.80  &     -0.36  &      0.91  &          - &          3 &    186.75  &    0.0004  &     -0.01  &      0.69  &     -0.05  &      0.66  &      0.05  &     -0.33  &      0.92  &     0.052  \\

 dT-0.029) &          4 &    180.36  &    0.0802  &     -0.01  &      0.01  &      1.00  &      0.00  &      0.99  &     -0.01  &          - &          4 &    179.74  &    0.1146  &     -0.02  &      0.04  &      1.00  &      0.02  &      0.00  &      0.98  &      0.00  &     0.012  \\

           &          5 &    171.51  &    0.2382  &      1.00  &      0.01  &      0.01  &      0.00  &      0.01  &      0.99  &          - &          5 &    177.08  &    0.2204  &      0.98  &      0.02  &      0.02  &      0.00  &      0.02  &     -0.01  &      0.97  &     0.011  \\

           &          6 &    160.85  &    0.0001  &      0.00  &      0.27  &     -0.01  &      0.95  &      0.94  &      0.31  &          - &          6 &    164.55  &    0.0009  &      0.03  &      0.21  &     -0.05  &      0.79  &      0.28  &      0.96  &      0.16  &     0.067  \\

           &          8 &    156.00  &    0.0003  &     -0.01  &      0.83  &      0.00  &     -0.53  &     -0.27  &      0.92  &          - &          7 &    156.79  &    0.0003  &     -0.02  &      0.84  &     -0.02  &     -0.48  &      0.04  &     -0.35  &      0.90  &     0.029  \\
\hline
       TD- &          1 &    249.48  &    0.0002  &     -0.04  &      0.97  &      0.00  &     -0.25  &      0.94  &      0.27  &     0.012  &          1 &    247.05  &    0.0000  &     -0.01  &      0.96  &      0.00  &     -0.25  &      0.01  &      0.94  &      0.28  &     0.026  \\

      CAM- &          2 &    233.87  &    0.1791  &      0.99  &      0.04  &     -0.08  &      0.00  &      1.00  &     -0.03  &     0.028  &          2 &    239.24  &    0.2568  &      0.98  &      0.01  &     -0.06  &      0.00  &      0.03  &      0.99  &     -0.02  &     0.024  \\

     B3LYP &          3 &    195.27  &    0.0000  &      0.00  &      0.60  &      0.00  &      0.79  &     -0.46  &      0.86  &     0.033  &          3 &    194.94  &    0.0004  &      0.04  &      0.60  &      0.00  &      0.76  &      0.05  &     -0.47  &      0.85  &     0.063  \\

      ($\sigma_{M}$: &          4 &    187.14  &    0.0683  &      0.08  &      0.00  &      1.00  &      0.00  &      1.00  &     -0.04  &     0.028  &          4 &    186.50  &    0.1413  &      0.97  &     -0.02  &      0.10  &     -0.03  &      0.03  &     -0.09  &      0.98  &     0.036  \\

Thy-0.027; &          5 &    180.42  &    0.1592  &      1.00  &      0.01  &      0.08  &      0.00  &      0.08  &      1.00  &     0.029  &            &            &            &     -0.10  &     -0.03  &      0.99  &     -0.01  &      0.01  &      0.98  &      0.08  &            \\

 dT-0.048) &          6 &    176.17  &    0.0000  &      0.00  &      0.28  &      0.00  &      0.95  &      0.91  &      0.42  &     0.033  &          5 &    184.44  &    0.1035  &      0.16  &      0.00  &      0.97  &      0.01  &      0.01  &      0.97  &      0.16  &     0.071  \\

           &          7 &    166.72  &    0.0002  &     -0.01  &      0.78  &      0.00  &     -0.62  &     -0.28  &      0.94  &     0.030  &          6 &    180.74  &    0.0029  &      0.09  &      0.19  &      0.00  &      0.73  &      0.36  &      0.91  &      0.35  &     0.064  \\

           &          9 &    155.18  &    0.3904  &     -0.08  &      0.00  &      0.99  &      0.00  &     -0.01  &      1.00  &          - &          7 &    169.05  &    0.0067  &     -0.10  &      0.09  &     -0.03  &      0.61  &      0.53  &      0.98  &     -0.01  &          - \\

           &            &            &            &            &            &            &            &            &            &            &          8 &    166.90  &    0.0014  &      0.00  &      0.77  &     -0.01  &     -0.61  &      0.02  &     -0.26  &      0.94  &     0.052  \\
\hline
       TD- &          1 &    249.72  &    0.0002  &     -0.04  &      0.97  &      0.00  &     -0.25  &      0.95  &      0.26  &     0.013  &          1 &    247.18  &    0.0000  &     -0.01  &      0.97  &      0.00  &     -0.25  &      0.00  &      0.94  &      0.27  &     0.024  \\

   $\omega$B97X-D &          2 &    234.05  &    0.1764  &      0.99  &      0.04  &     -0.08  &      0.00  &      1.00  &     -0.03  &     0.028  &          2 &    239.44  &    0.2531  &      0.98  &      0.01  &     -0.06  &      0.01  &      0.03  &      0.99  &     -0.02  &     0.024  \\

      ($\sigma_{M}$: &          3 &    194.84  &    0.0000  &     -0.02  &      0.61  &      0.00  &      0.78  &     -0.49  &      0.85  &     0.042  &          3 &    194.60  &    0.0004  &      0.04  &      0.60  &      0.00  &      0.75  &      0.05  &     -0.50  &      0.84  &     0.069  \\

Thy-0.032; &          4 &    188.03  &    0.0673  &      0.08  &      0.00  &      0.99  &      0.00  &      1.00  &     -0.05  &     0.029  &          4 &    187.12  &    0.1232  &      0.89  &     -0.03  &      0.41  &     -0.03  &      0.03  &     -0.40  &      0.90  &     0.163  \\

 dT-0.088) &          5 &    180.66  &    0.1580  &      0.99  &      0.01  &      0.09  &      0.00  &      0.08  &      1.00  &     0.031  &            &            &            &     -0.41  &     -0.01  &      0.91  &      0.01  &      0.01  &      0.90  &      0.39  &            \\

           &          6 &    177.27  &    0.0001  &      0.00  &      0.29  &      0.00  &      0.94  &      0.90  &      0.43  &     0.037  &          5 &    184.84  &    0.1186  &      0.46  &     -0.01  &      0.86  &      0.00  &      0.03  &      0.86  &      0.47  &     0.202  \\

           &          7 &    167.82  &    0.0002  &      0.00  &      0.76  &      0.00  &     -0.65  &     -0.25  &      0.95  &     0.042  &            &            &            &      0.86  &     -0.02  &     -0.48  &     -0.02  &      0.02  &     -0.49  &      0.86  &            \\

           &          9 &    155.06  &    0.4063  &     -0.08  &      0.00  &      0.99  &      0.00  &     -0.02  &      1.00  &          - &          6 &    181.44  &    0.0024  &      0.09  &      0.21  &      0.01  &      0.75  &      0.31  &      0.90  &      0.38  &     0.071  \\

           &            &            &            &            &            &            &            &            &            &            &          7 &    169.70  &    0.0064  &     -0.10  &      0.05  &     -0.04  &      0.58  &      0.56  &      0.98  &     -0.06  &          - \\

           &            &            &            &            &            &            &            &            &            &            &          8 &    167.85  &    0.0018  &      0.00  &      0.75  &     -0.01  &     -0.64  &      0.02  &     -0.23  &      0.95  &     0.065  \\
\hline
       TD- &          1 &    242.13  &    0.0002  &     -0.05  &      0.96  &      0.00  &     -0.28  &      0.92  &      0.31  &     0.021  &          1 &    239.80  &    0.0000  &     -0.01  &      0.96  &      0.00  &     -0.28  &      0.00  &      0.92  &      0.32  &     0.041  \\

     $\omega$B97X &          2 &    228.82  &    0.2013  &      1.00  &      0.05  &     -0.04  &      0.00  &      1.00  &     -0.05  &     0.022  &          2 &    233.96  &    0.2820  &      0.98  &      0.01  &     -0.02  &      0.00  &      0.03  &      0.99  &     -0.03  &     0.015  \\

      ($\sigma_{M}$: &          3 &    190.78  &    0.0000  &      0.00  &      0.57  &      0.00  &      0.81  &     -0.35  &      0.90  &     0.009  &          3 &    190.28  &    0.0003  &      0.03  &      0.57  &     -0.01  &      0.79  &      0.03  &     -0.34  &      0.90  &     0.054  \\

Thy-0.018; &          4 &    178.49  &    0.0659  &      0.05  &      0.00  &      0.99  &      0.00  &      1.00  &      0.06  &     0.027  &          4 &    179.36  &    0.1967  &      0.98  &     -0.01  &      0.04  &     -0.02  &      0.03  &      0.04  &      0.98  &     0.019  \\

 dT-0.035) &          5 &    174.43  &    0.1980  &      1.00  &      0.01  &      0.03  &      0.00  &      0.06  &      1.00  &     0.016  &          5 &    176.86  &    0.0825  &      0.02  &     -0.01  &      0.99  &      0.00  &      0.00  &      0.99  &      0.08  &     0.030  \\

           &          6 &    164.14  &    0.0002  &      0.00  &      0.26  &      0.00  &      0.94  &      0.95  &      0.30  &     0.006  &          6 &    168.19  &    0.0015  &      0.05  &      0.20  &      0.00  &      0.74  &      0.29  &      0.95  &      0.24  &     0.032  \\

           &          7 &    158.49  &    0.0003  &     -0.01  &      0.83  &      0.00  &     -0.55  &     -0.34  &      0.90  &     0.022  &          7 &    159.11  &    0.0010  &      0.00  &      0.78  &      0.02  &     -0.55  &      0.03  &     -0.48  &      0.84  &     0.051  \\

           &          9 &    151.73  &    0.3700  &     -0.05  &      0.00  &      0.99  &      0.00  &     -0.08  &      1.00  &          - &          8 &    156.42  &    0.0086  &     -0.08  &      0.28  &     -0.05  &      0.50  &      0.54  &      0.97  &      0.06  &          - \\
\hline
       TD- &          1 &    238.63  &    0.0002  &     -0.05  &      0.95  &      0.00  &     -0.29  &      0.91  &      0.33  &     0.027  &          1 &    236.44  &    0.0000  &     -0.01  &      0.95  &      0.00  &     -0.29  &      0.00  &      0.91  &      0.34  &     0.048  \\

      $\omega$B97 &          2 &    226.03  &    0.2140  &      1.00  &      0.05  &     -0.02  &      0.00  &      1.00  &     -0.06  &     0.021  &          2 &    231.07  &    0.2966  &      0.98  &      0.02  &     -0.01  &      0.00  &      0.03  &      0.99  &     -0.04  &     0.015  \\

      ($\sigma_{M}$: &          3 &    189.05  &    0.0000  &      0.00  &      0.55  &      0.00  &      0.83  &     -0.31  &      0.91  &     0.021  &          3 &    188.48  &    0.0002  &      0.03  &      0.55  &     -0.01  &      0.81  &      0.03  &     -0.31  &      0.91  &     0.062  \\

Thy-0.028; &          4 &    173.59  &    0.0642  &      0.08  &      0.00  &      0.99  &      0.00  &      0.99  &      0.12  &     0.046  &          4 &    175.51  &    0.1967  &      0.97  &     -0.01  &      0.05  &     -0.02  &      0.03  &      0.07  &      0.98  &     0.027  \\

 dT-0.040) &          5 &    170.20  &    0.2295  &      1.00  &      0.01  &     -0.04  &      0.00  &      0.06  &      1.00  &     0.021  &          5 &    171.75  &    0.1108  &     -0.03  &     -0.01  &      0.99  &      0.00  &      0.00  &      0.98  &      0.14  &     0.043  \\

           &          6 &    158.03  &    0.0004  &      0.00  &      0.24  &      0.00  &      0.91  &      0.97  &      0.22  &     0.031  &          6 &    162.13  &    0.0016  &      0.04  &      0.20  &      0.01  &      0.72  &      0.29  &      0.96  &      0.19  &     0.028  \\

           &          7 &    153.85  &    0.0004  &     -0.01  &      0.85  &      0.00  &     -0.49  &     -0.36  &      0.89  &     0.030  &          7 &    154.77  &    0.0013  &     -0.01  &      0.79  &      0.02  &     -0.50  &      0.03  &     -0.51  &      0.81  &     0.056  \\

           &          9 &    149.65  &    0.3419  &     -0.03  &      0.00  &      0.99  &      0.00  &     -0.14  &      0.99  &          - &          8 &    150.79  &    0.0208  &     -0.05  &      0.30  &     -0.09  &      0.49  &      0.51  &      0.97  &      0.04  &          - \\
\hline
       TD- &          1 &    241.89  &    0.0002  &     -0.04  &      0.95  &      0.00  &     -0.29  &      0.91  &      0.34  &     0.028  &          1 &    239.42  &    0.0000  &     -0.01  &      0.95  &      0.00  &     -0.29  &      0.00  &      0.90  &      0.34  &     0.048  \\

   LC-$\omega$PBE &          2 &    226.50  &    0.2118  &      1.00  &      0.04  &     -0.02  &      0.00  &      1.00  &     -0.06  &     0.019  &          2 &    231.30  &    0.2938  &      0.98  &      0.02  &     -0.01  &      0.00  &      0.03  &      0.99  &     -0.04  &     0.015  \\

      ($\sigma_{M}$: &          3 &    191.81  &    0.0000  &      0.00  &      0.54  &      0.00  &      0.83  &     -0.31  &      0.91  &     0.023  &          3 &    191.01  &    0.0002  &      0.03  &      0.54  &     -0.01  &      0.82  &      0.03  &     -0.30  &      0.91  &     0.066  \\

Thy-0.029; &          4 &    174.37  &    0.0663  &      0.09  &      0.00  &      0.99  &      0.00  &      0.99  &      0.13  &     0.050  &          4 &    176.69  &    0.2022  &      0.97  &      0.01  &      0.05  &     -0.02  &      0.03  &      0.06  &      0.98  &     0.023  \\

 dT-0.040) &          5 &    171.41  &    0.2327  &      1.00  &      0.01  &     -0.04  &      0.00  &      0.06  &      1.00  &     0.021  &          5 &    172.37  &    0.1125  &     -0.02  &     -0.01  &      0.99  &      0.00  &      0.00  &      0.98  &      0.15  &     0.046  \\

           &          6 &    159.28  &    0.0004  &      0.00  &      0.24  &      0.00  &      0.91  &      0.97  &      0.22  &     0.031  &          6 &    163.08  &    0.0014  &      0.04  &      0.20  &      0.01  &      0.72  &      0.28  &      0.96  &      0.19  &     0.028  \\

           &          7 &    155.34  &    0.0004  &     -0.01  &      0.85  &      0.00  &     -0.49  &     -0.37  &      0.88  &     0.034  &          7 &    156.01  &    0.0010  &     -0.01  &      0.81  &      0.02  &     -0.49  &      0.03  &     -0.50  &      0.82  &     0.051  \\

           &          8 &    150.72  &    0.3484  &     -0.03  &      0.00  &      0.99  &      0.00  &     -0.14  &      0.99  &          - &          8 &    151.61  &    0.0317  &     -0.04  &      0.29  &     -0.12  &      0.49  &      0.51  &      0.97  &      0.03  &          - \\
\hline
\end{tabular}
}
\end{table*}

\begin{table*}[t]
\raggedright

{\fontsize{6pt}{12pt}\selectfont
\begin{tabular}{|c|ccc|cccc|cc|c||ccc|ccccc|cc|c|}
\hline
     (b) &  \multicolumn{ 3}{c}{Thy} &                   \multicolumn{ 4}{c}{NTO1(2)-H} & \multicolumn{ 2}{c}{NTO1(2)-E} &           &  \multicolumn{ 3}{c}{dT} &                                \multicolumn{ 5}{c}{NTO1(2)-H} & \multicolumn{ 2}{c}{NTO1(2)-E} &           \\

  &          N &      $\lambda$(nm) &       f &         P1 &         N1 &         P2 &         N2 &         S1 &         S2 &   $\sigma$         &          N &      $\lambda$(nm) &       f &         P1 &         N1 &         P2 &         N2 &       (B)$^{2}$ &         S1 &         S2 &   $\sigma$         \\
\hline
       CIS &          1 &    197.73  &    0.0077  &     -0.13  &      0.93  &     -0.01  &     -0.33  &      0.86  &      0.40  &     0.062  &          1 &    198.84  &    0.5502  &      0.98  &      0.01  &      0.03  &      0.00  &      0.02  &      0.98  &     -0.06  &     0.018  \\

      ($\sigma_{M}$: &          2 &    195.68  &    0.4332  &      0.98  &      0.12  &      0.02  &     -0.04  &      0.99  &     -0.07  &     0.038  &          2 &    195.34  &    0.0001  &     -0.02  &      0.94  &     -0.01  &     -0.32  &      0.00  &      0.85  &      0.41  &     0.074  \\

Thy-0.138; &          3 &    161.65  &    0.0000  &     -0.01  &      0.47  &     -0.01  &      0.87  &     -0.23  &      0.91  &     0.055  &          3 &    160.17  &    0.0005  &      0.03  &      0.48  &     -0.02  &      0.86  &      0.02  &      0.23  &     -0.91  &     0.090  \\

 dT-0.080) &          4 &    147.53  &    0.1508  &      0.80  &      0.02  &      0.58  &      0.00  &      0.56  &      0.80  &     0.242  &          4 &    149.55  &    0.2503  &      0.98  &      0.00  &      0.07  &     -0.03  &      0.03  &     -0.10  &     -0.98  &     0.037  \\

           &            &            &            &     -0.59  &      0.00  &      0.79  &      0.00  &      0.79  &     -0.58  &            &          5 &    141.28  &    0.2894  &     -0.21  &      0.00  &      0.96  &      0.01  &      0.00  &      0.85  &      0.49  &     0.157  \\

           &          6 &    141.31  &    0.4179  &     -0.61  &      0.00  &      0.78  &      0.01  &      0.62  &      0.77  &     0.310  &          7 &    126.31  &    0.2175  &      0.10  &      0.03  &      0.97  &     -0.01  &      0.00  &      0.41  &     -0.90  &          - \\

           &          8 &    126.65  &    0.2298  &      0.13  &      0.01  &      0.97  &      0.00  &     -0.38  &      0.92  &          - &         15 &    115.65  &    0.0033  &      0.04  &     -0.46  &      0.01  &     -0.56  &      0.07  &      0.96  &      0.03  &     0.108  \\

           &         12 &    116.52  &    0.0004  &      0.00  &      0.59  &     -0.01  &      0.61  &      0.98  &      0.01  &          - &            &            &            &      0.90  &      0.09  &      0.14  &      0.02  &      0.13  &     -0.08  &      0.05  &            \\

           &            &            &            &     -0.01  &      0.63  &      0.00  &     -0.66  &      0.05  &      0.93  &            &            &           &            &            &            &            &            &            &            &            &            \\

           &         15 &    110.63  &    0.0001  &      0.00  &      0.03  &      0.00  &      0.63  &      0.98  &      0.17  &     0.123  &            &            &            &            &            &            &            &            &            &            &            \\

           &            &            &            &     -0.02  &      0.69  &      0.00  &     -0.40  &     -0.19  &      0.92  &            &            &            &            &            &            &            &            &            &            &            &            \\
\hline
     TD-HF &          1 &    206.52  &    0.3427  &      0.98  &     -0.04  &      0.04  &      0.02  &      0.99  &     -0.07  &     0.025  &          1 &    209.49  &    0.4576  &      0.97  &      0.01  &      0.04  &      0.00  &      0.01  &      0.98  &     -0.06  &     0.019  \\

      ($\sigma_{M}$: &          2 &    202.72  &    0.0021  &      0.05  &      0.94  &      0.00  &     -0.33  &      0.85  &      0.41  &     0.057  &          2 &    200.20  &    0.0000  &     -0.02  &      0.94  &     -0.01  &     -0.32  &      0.00  &      0.84  &      0.41  &     0.075  \\

Thy-0.137; &          3 &    164.91  &    0.0000  &     -0.01  &      0.47  &     -0.01  &      0.87  &     -0.23  &      0.91  &     0.055  &          3 &    163.31  &    0.0005  &      0.04  &      0.48  &     -0.02  &      0.86  &      0.02  &      0.23  &     -0.91  &     0.090  \\

 dT-0.106) &          4 &    152.36  &    0.1127  &      0.65  &      0.01  &      0.73  &      0.00  &      0.70  &      0.67  &     0.297  &          4 &    153.99  &    0.1954  &      0.97  &      0.00  &      0.09  &     -0.04  &      0.02  &     -0.11  &     -0.97  &     0.042  \\

           &            &            &            &     -0.75  &     -0.01  &      0.64  &      0.00  &      0.67  &     -0.73  &            &          5 &    144.95  &    0.2422  &     -0.31  &      0.01  &      0.93  &      0.02  &      0.00  &      0.79  &      0.59  &     0.199  \\

           &          5 &    144.72  &    0.3622  &      0.66  &      0.00  &     -0.74  &      0.00  &      0.53  &      0.84  &     0.284  &          7 &    128.95  &    0.1427  &      0.14  &      0.04  &      0.96  &     -0.01  &      0.00  &      0.43  &     -0.89  &          - \\

           &          8 &    129.32  &    0.1426  &      0.18  &      0.03  &      0.95  &     -0.01  &     -0.41  &      0.90  &          - &         15 &    116.43  &    0.0077  &      0.50  &     -0.30  &      0.07  &     -0.39  &      0.28  &      0.69  &      0.06  &     0.211  \\

           &         15 &    111.24  &    0.0004  &      0.00  &      0.08  &      0.00  &      0.65  &      0.97  &      0.21  &     0.107  &            &            &            &      0.70  &      0.32  &      0.08  &      0.34  &      0.04  &      0.64  &     -0.02  &            \\

           &         22 &    100.71  &    0.0099  &      0.61  &      0.06  &      0.05  &      0.00  &     -0.13  &      0.70  &          - &         21 &    110.25  &    0.0317  &      0.96  &      0.03  &      0.01  &      0.00  &      0.02  &      0.87  &      0.25  &          - \\

           &            &            &            &     -0.02  &      0.88  &     -0.01  &      0.44  &      0.01  &     -0.04  &            &            &            &            &            &            &            &            &            &            &            &            \\
\hline
       TD- &          1 &    270.56  &    0.0001  &     -0.04  &      0.99  &      0.00  &     -0.15  &      0.99  &      0.15  &     0.048  &          1 &    268.35  &    0.0001  &     -0.03  &      0.98  &      0.00  &     -0.15  &      0.01  &      0.97  &      0.16  &     0.026  \\

     B3LYP &          2 &    246.47  &    0.1268  &      0.97  &      0.04  &     -0.21  &      0.00  &      1.00  &      0.00  &     0.064  &          2 &    253.13  &    0.1897  &      0.94  &      0.02  &     -0.16  &     -0.03  &      0.07  &      0.99  &      0.01  &     0.054  \\

      ($\sigma_{M}$: &          3 &    216.14  &    0.0000  &      0.00  &      0.25  &      0.00  &      0.96  &      0.97  &     -0.24  &     0.159  &          3 &    220.32  &    0.0042  &      0.15  &      0.11  &     -0.02  &      0.74  &      0.37  &      0.98  &     -0.10  &     0.089  \\

Thy-0.116; &          4 &    208.21  &    0.0574  &      0.23  &      0.00  &      0.97  &      0.00  &      0.97  &     -0.22  &     0.093  &          4 &    207.29  &    0.0591  &      0.26  &     -0.03  &      0.96  &     -0.04  &      0.01  &      0.95  &     -0.28  &     0.118  \\

 dT-0.110) &          5 &    200.48  &    0.0001  &     -0.05  &      0.91  &      0.00  &      0.40  &      0.10  &      0.99  &     0.200  &          5 &    205.43  &    0.0003  &     -0.16  &      0.53  &      0.12  &      0.53  &      0.39  &     -0.87  &      0.47  &     0.221  \\

           &          6 &    192.17  &    0.1223  &      0.99  &      0.04  &      0.15  &      0.05  &      0.14  &      0.99  &     0.058  &            &            &            &      0.01  &      0.69  &     -0.15  &      0.12  &      0.45  &      0.46  &      0.86  &            \\

           &          7 &    184.80  &    0.0002  &      0.03  &      0.38  &      0.00  &     -0.92  &     -0.03  &      1.00  &     0.187  &          6 &    200.39  &    0.0014  &      0.08  &      0.67  &      0.11  &      0.68  &      0.07  &      0.60  &      0.77  &          - \\

           &          9 &    162.81  &    0.3352  &     -0.16  &      0.00  &      0.98  &      0.00  &      0.01  &      1.00  &          - &            &            &            &      0.21  &      0.62  &     -0.07  &     -0.43  &      0.35  &     -0.78  &      0.61  &            \\

           &            &            &            &            &            &            &            &            &            &            &          7 &    197.43  &    0.1530  &      0.96  &     -0.04  &      0.18  &     -0.04  &      0.04  &      0.19  &      0.95  &     0.077  \\

           &            &            &            &            &            &            &            &            &            &            &          9 &    186.15  &    0.0003  &     -0.04  &      0.42  &      0.00  &     -0.87  &      0.06  &     -0.07  &      0.98  &     0.186  \\
\hline
       TD- &          1 &    263.26  &    0.0001  &     -0.04  &      0.98  &      0.00  &     -0.18  &      0.98  &      0.18  &     0.036  &          1 &    261.13  &    0.0001  &     -0.03  &      0.98  &      0.00  &     -0.18  &      0.01  &      0.97  &      0.19  &     0.017  \\

   PBE1PBE &          2 &    241.45  &    0.1414  &      0.98  &      0.04  &     -0.17  &      0.00  &      1.00  &     -0.01  &     0.053  &          2 &    247.82  &    0.2086  &      0.95  &      0.02  &     -0.13  &     -0.02  &      0.06  &      0.99  &      0.00  &     0.045  \\

      ($\sigma_{M}$: &          3 &    207.93  &    0.0000  &     -0.01  &      0.37  &      0.00  &      0.92  &      0.91  &     -0.41  &     0.210  &          3 &    211.64  &    0.0033  &      0.13  &      0.17  &     -0.01  &      0.74  &      0.37  &      0.97  &     -0.17  &     0.102  \\

Thy-0.113; &          4 &    201.76  &    0.0589  &      0.19  &      0.00  &      0.98  &      0.00  &      0.98  &     -0.18  &     0.076  &          4 &    200.76  &    0.0643  &      0.22  &     -0.03  &      0.97  &     -0.03  &      0.01  &      0.96  &     -0.24  &     0.102  \\

 dT-0.096) &          5 &    193.99  &    0.0002  &     -0.05  &      0.79  &      0.00  &      0.61  &      0.32  &      0.95  &     0.212  &          5 &    199.04  &    0.0001  &     -0.11  &      0.82  &      0.02  &      0.54  &      0.03  &     -0.37  &      0.92  &     0.063  \\

           &          6 &    187.83  &    0.1333  &      0.99  &      0.04  &      0.13  &      0.05  &      0.12  &      0.99  &     0.050  &            &            &            &     -0.10  &     -0.13  &      0.19  &      0.02  &      0.89  &      0.91  &      0.35  &            \\

           &          7 &    179.67  &    0.0002  &      0.02  &      0.49  &      0.00  &     -0.87  &     -0.08  &      1.00  &     0.151  &          6 &    193.41  &    0.0911  &      0.79  &      0.38  &      0.23  &      0.35  &      0.05  &      0.32  &      0.92  &     0.192  \\

           &          9 &    159.34  &    0.4077  &     -0.13  &      0.00  &      0.99  &      0.00  &      0.02  &      1.00  &          - &            &            &            &     -0.40  &     -0.17  &      0.61  &      0.52  &      0.14  &      0.93  &     -0.33  &            \\

           &            &            &            &            &            &            &            &            &            &            &          7 &    192.11  &    0.0688  &     -0.41  &      0.29  &     -0.28  &      0.69  &      0.20  &      0.87  &      0.45  &          - \\

           &            &            &            &            &            &            &            &            &            &            &            &            &            &      0.63  &     -0.47  &     -0.15  &      0.17  &      0.31  &     -0.46  &      0.87  &            \\

           &            &            &            &            &            &            &            &            &            &            &          9 &    180.60  &    0.0003  &     -0.02  &      0.53  &      0.01  &     -0.82  &      0.04  &     -0.11  &      0.98  &     0.152  \\
\hline
       TD- &          1 &    316.41  &    0.0000  &     -0.03  &      1.00  &      0.00  &      0.05  &      1.00  &     -0.04  &     0.125  &          1 &    316.27  &    0.0006  &     -0.07  &      0.95  &      0.02  &      0.07  &      0.07  &      0.99  &     -0.02  &          - \\

      LDA &          2 &    270.51  &    0.0662  &      0.90  &      0.03  &     -0.40  &      0.05  &      0.99  &      0.08  &     0.125  &          2 &    291.65  &    0.0413  &      0.50  &      0.27  &     -0.16  &     -0.28  &      0.53  &      0.99  &      0.01  &          - \\

     ($\sigma_{M}$ : &          3 &    267.18  &    0.0003  &     -0.05  &     -0.05  &      0.01  &      0.99  &      1.00  &     -0.01  &     0.133  &          3 &    271.79  &    0.0622  &      0.67  &     -0.08  &     -0.25  &      0.59  &      0.11  &      0.98  &      0.05  &          - \\

Thy-0.201) &          4 &    236.19  &    0.0011  &     -0.08  &      0.99  &     -0.03  &     -0.07  &      0.04  &      1.00  &     0.170  &          5 &    254.91  &    0.0125  &     -0.33  &      0.07  &      0.14  &      0.73  &      0.34  &      0.99  &      0.03  &          - \\

           &          5 &    235.14  &    0.0507  &      0.43  &      0.06  &      0.90  &      0.01  &      0.91  &     -0.40  &     0.174  &          6 &    239.34  &    0.0006  &     -0.08  &      0.97  &     -0.02  &     -0.05  &      0.04  &      0.01  &      0.98  &          - \\

           &          6 &    208.80  &    0.0944  &      0.97  &      0.04  &      0.20  &      0.02  &      0.22  &      0.97  &     0.083  &          7 &    235.80  &    0.0456  &      0.40  &      0.05  &      0.86  &     -0.03  &      0.09  &      0.90  &     -0.41  &          - \\

           &          7 &    203.29  &    0.0001  &     -0.02  &      0.06  &     -0.01  &      0.99  &      0.00  &      1.00  &     0.195  &          9 &    220.59  &    0.0679  &      0.42  &      0.15  &      0.03  &     -0.33  &      0.65  &      0.02  &      0.97  &          - \\

           &         11 &    173.87  &    0.2857  &     -0.19  &      0.00  &      0.97  &      0.01  &     -0.05  &      1.00  &          - &         10 &    214.82  &    0.0548  &      0.78  &     -0.05  &      0.15  &      0.35  &      0.22  &      0.11  &      0.93  &          - \\

           &            &            &            &      0.34  &     -0.01  &      0.16  &      0.00  &      1.00  &      0.05  &            &            &            &            &            &            &            &            &            &            &            &            \\
\hline
\end{tabular}
}
\end{table*}

\end{document}